\documentclass[preprint,review,12pt]{elsarticle}

\usepackage{amssymb}
\usepackage[justification=centering]{caption}
\usepackage{rotating}
\usepackage{multirow}
\usepackage{bm}
\usepackage{tabularx,booktabs}
\usepackage[T1]{fontenc}
\usepackage{blindtext}
\usepackage{amsbsy}
\usepackage{amsfonts}
\usepackage{amsmath,tabstackengine}           % to use \boldsymbol{}, \bmatrix{}
\usepackage[]{natbib}
\usepackage{epstopdf}
\usepackage{float}
\usepackage{placeins}
\usepackage{enumitem}
\usepackage[usenames,dvipsnames]{xcolor}
\usepackage[bookmarks,bookmarksnumbered,colorlinks]{hyperref}
\usepackage{graphicx}
\usepackage{geometry}
\usepackage{arydshln}  %%% to produce dashed lines
\AtBeginDocument{\hypersetup{citecolor=blue,urlcolor=brown,linkcolor=Blue}}
\usepackage[font=normalsize,labelfont=bf]{caption}
\usepackage{subcaption}
\usepackage{placeins}
\usepackage{lipsum}
\DeclareCaptionFormat{cont}{#1#2#3\par}

\newcommand{\itwas}[1]{{\color{red} #1}}

\newcommand\degr{^{\circ}}

\biboptions{authoryear}

\journal{International Journal of Solids and Structures}

\begin{document}

\begin{frontmatter}

\title{
Cylindrical void growth vs. grain fragmentation in FCC single crystals: CPFEM study for two types of loading conditions}

\author[inst1]{Saketh Virupakshi}

\affiliation[inst1]{organization={Institute of Fundamental Technological Research, Polish Academy of Sciences},%Department and Organization
            addressline={Pawinskiego 5b}, 
            city={Warsaw},
            postcode={02106}, 
            country={Poland}}

\author[inst1]{Katarzyna Kowalczyk-Gajewska \corref{cor1}}
\cortext[cor1]{Corresponding author.}
\ead{kkowalcz@ippt.pan.pl}
\begin{abstract}
%% Text of abstract
The crystal plasticity finite element method (CPFEM) is used to investigate the coupling between the cylindrical void growth or collapse and grain refinement in face-centered cubic (FCC) single crystals. A 2D plane strain model with one void is used. The effect of the initial lattice orientation, similarities, and differences between stress- and strain-driven loading scenarios are explored. To this end, boundary conditions are enforced in two different ways. The first one is based on maintaining constant in-plane stress biaxiality via a dedicated truss element, while the second one is imposing a constant displacement biaxiality factor. Uniaxial and biaxial loading cases are studied. For the uniaxial loading case a special configuration, which enforces an equivalent pattern of plastic deformation in the pristine crystal, is selected in order to investigate the mutual interactions between the evolving void and the developed lattice rotation heterogeneity. Next, biaxial loading cases are considered for three crystal orientations, one of which is not symmetric with respect to loading directions. It is analysed how stress or strain biaxility factors and initial lattice orientation influence the void evolution in terms of its size and shape. Moreover, the consequences of variations in the resulting heterogeneity of lattice rotation are studied in the context of the grain refinement phenomenon accompanying the void evolution. Scenarios that may lead to more advanced grain fragmentation are identified.
\end{abstract}


\end{frontmatter}
%% main text
\section{Introduction}\label{sec:Introduction} 

Nucleation, growth, and coalescence of intra- or intergranular micro-voids is a usual scenario by which ductile metallic materials fail \citep{BenzergaLeblond10}. Most often, micro-voids are nucleated as a result of decohesion or fracture process of second phase precipitates. Growth of those micro-defects takes place due to diffuse plastic deformation up to the onset of coalescence when strain localizes in the ligament connecting closely spaced voids. From the mechanics point of view, for 3D stress controlled axially symmetric loading (e.g. uniaxial tension process) the void coalescence is connected with the transition from axisymmetric to uniaxial straining mode, which leads to the plastic flow localization.
%From the mechanics point of view, the void coalescence is connected with the transition from axisymmetric to uniaxial straining mode, which leads to the plastic flow localization.  
After that moment the voids continue expansion, mostly towards each other up to final ligament failure or full impingement. On the other hand, the development of strain heterogeneity around a deforming and growing void leads to microstructure changes in the material around the void. It seems that the last aspect of the mechanics of porous metallic materials has not been fully explored, yet. On the other hand, this effect accompanying void evolution can have important consequences as concerns the grain refinement as observed by \cite{Beygelzimer05} who formulated a phenomenological model of grain fragmentation.

Beginning from the late 60s of the previous century a lot of studies have been done experimentally, theoretically, and numerically to understand the mechanics of void initiation, growth, and coalescence. Initially, numerical analyses were performed using the macroscopic isotropic nearly rate-independent elastic-plastic model of metallic material. \cite{Tvergaard82} and \cite{Koplik88} seem to be the first who applied respectively the 2D and 3D unit cell approach in this respect. Based on performed studies the Gurson yield criterion for porous metals, originally developed based on the analytical solution and micromechanical approach, has been modified and equipped with additional tunning parameters to become the widely used GTN (Gurson-Tvergaard-Needleman) criterion \citep{Gurson77,Tvergaard82,Tvergaard84}. Those initial studies revealed an important role of stress triaxiality in the void growth phenomenon. 
In the next studies, the influence of the third invariant (Lode parameter) and material anisotropy \citep{Benzerga01} was observed. However, as recently concluded by \cite{Pineau16} there is not yet a universal theory of ductile failure. Moreover, as expressed by a recent review by \cite{Das21} there is no  full agreement yet if the process is more strain or stress-driven, while these two mechanical fields are strictly related by a constitutive behaviour of the virgin medium and problem geometry (i.e. void shape and space distribution).

The void growth and coalescence contributing to the ductile failure is a process that usually takes place at the micro-level of polycrystalline metal, therefore it was soon understood that replacing the macroscopic plasticity model for a matrix material in numerical calculations with a more relevant continuum crystal plasticity framework may lead to a better understanding of the phenomenon.  Following this observation, firstly, 2D analyses of unit cells with cylindrical voids were initiated under plane strain and in-plane strain-driven boundary conditions \citep{Oregan97,POTIRNICHE2006921}.  
Similar 3D analyses of spherical void growth and coalescence under strain-controlled boundary conditions were performed by \cite{Liu07}. Those boundary conditions (i.e. strain-based) were motivated by the possibility of avoiding any instability in the calculations. The main conclusion of those studies was that the influence of crystal orientation is more significant for the loading cases with a small strain biaxility or triaxility and of secondary importance for higher strain biaxiality or triaxiality factors. Among mentioned studies, only \cite{Liu07} provided some results related to microstructure evolution in the presence of voids. They are concerned with the texture evolution within the unit cell and the heterogeneity of deformation assessed by the misorientation angle with respect to the average orientation. It was concluded that the heterogeneity of lattice rotation is concentrated in the regions around the void. 

The cylindrical void growth under plane strain while constant in-plane stress biaxiality factor in hexagonal close packed (HCP) crystal studied by \cite{SubrahmanyaPrasad2015NumericalSO}. \cite{Yerra10}
analysed 3D cell with a spherical void maintaining constant stress triaxiality, however applied boundary conditions were not purely stress driven since at the same time equal values of two lateral macroscopic strains were imposed. The macroscopic stress direction was fully controlled in the 3D calculations by \cite{SRIVASTAVA20131169,SRIVASTAVA201510} for Ni FCC single crystal, \cite{LING201658} for FCC austenitic stainless steel, and by \cite{SELVARAJOU2019198} for HCP Mg crystal. Different values of stress triaxiality and Lode parameter as well as selected crystal orientations with respect to loading axes were analysed.  
It was found that the value of the Lode parameter is more decisive concerning the void coalescence or collapse at a lower value of stress triaxiality. 
However, for certain anisotropic orientations, the Lode parameter can also have a significant effect on creep strain and porous evolution at higher stress triaxiality values \citep{SRIVASTAVA201510}, which led to forming a polygonal void shape with rounded corners. Moreover, the initial crystal orientation dictates the location of maximum stress concentration.  Based on such numerical studies analogues of the GTN criterion for single crystal were formulated by \cite{HAN20132115} and \cite{PAUX20151}, using the classical multi-surface Schmid condition and the regularized Schmid law, respectively, as valid models for the bulk medium.

Let us also remark that the void growth and coalescence in the heterogeneous bulk medium described by the crystal plasticity constitutive model were also studied numerically. For example, bi-crystal unit cells were assumed in \citep{Liu10b,Jeong18,Dakshinamurthy21} and polycrystal unit cells in \citep{Lebensohn13,Liu21}. Among mentioned contributions, selected results concerning heterogeneity of lattice rotation were provided in \citep{Dakshinamurthy21}. 
Unless stated otherwise, a large strain rate-dependent CP formulation with the power law for slip was used in all papers recalled above. 

The goal of the present research is two-fold. First, we would like to compare and analyse strain and stress-driven loading scenarios in the context of cylindrical void evolution in FCC single crystal under plane strain conditions. In particular, the effect of strain vs. stress in-plane biaxiality factor is elucidated. To our best knowledge, such direct comparison has not been performed in the literature yet. Second, the mutual interactions between the void evolution and development of lattice rotation heterogeneity, leading to grain refinement, as two competitive mechanisms of microstructure changes are explored. Such an interplay between two effects seems not to be sufficiently quantified in other research.

The paper is organized as follows. After this introductory section, we present the applied crystal plasticity model and its finite element implementation in Section 2. Section 3 is devoted to the description of the numerical model of a unit cell and the boundary conditions. The main body of the paper is included in Section 4 where the results of performed numerical studies are outlined and discussed. Their presentation is divided into two parts. The first concerns four uniaxial loading cases under a special crystal configuration, which enable the observation of a clear coupling of the void growth or collapse with the grain fragmentation into subgrains. The second part of this section continues the analysis of the impact of two biaxiality factors on the void growth and overall stress-strain response for biaxial loading processes, as well as their relation to microstructure evolution. The paper is closed with conclusions.
\section{Crystal plasticity model and its FE implementation}
\subsection{Crystal plasticity constitutive theory}

In this section, the key details of crystal plasticity implementation in FEM applied in the analyses are described. Model formulation and implementation follow \cite{Kucharski14} and \cite{Frydrych_2018}.

{First, let us present the rate-dependent elastic-plastic model of the single crystal. In terms of kinematics description, the model follows classical contributions by \cite{HILL1972401,ASARO1977309,ASARO1985923}. The deformation gradient $\mathbf{F}$ is multiplicatively decomposed into two parts:}
	\begin{equation}  \label{eq:Multiplicative}
		\mathbf{F}=\mathbf{F}_e\mathbf{F}_p\
	\end{equation}
	where $\mathbf{F}_e$ and $\mathbf{F}_p$ denote the elastic and plastic components, respectively. The evolution of the plastic part of the deformation gradient is governed by the equation:
	\begin{equation}\label{Eq:Fpevolution}
		\dot{\mathbf{F}}_p=\hat{\mathbf{L}}_p\mathbf{F}_p,
	\end{equation}
	where the dot over the quantity denotes its material time derivative. The plastic velocity gradient $\hat{\mathbf{L}}_p$ is the sum of shears on slip systems:
	\begin{equation}
		\hat{\mathbf{L}}_p=\sum_{r=1}^{M}\dot{\gamma}^r\mathbf{m}^r_0\otimes\mathbf{n}^r_0
	\end{equation}
	{with unit vectors} $\mathbf{m}^r_0$ and $\mathbf{n}^r_0$ denoting the $r$-th slip system direction and plane normal defined in the initial configuration. In FCC crystals, plastic deformation occurs along the \{111\}$\langle110\rangle$ family of slip systems, which contains 12 potentially active slip systems that are taken into account in the computations.
	
	In the \emph{rate-dependent} formulation, in order to calculate shear rates the power law \citep{Asaro85} is used:

\begin{equation}
	\dot{\gamma}^r=v_0\mathrm{sign}(\tau^r)\left|\frac{\tau^r}{\tau_c^r}\right|^{\bar{n}}
\end{equation}

    \noindent{where $v_0$ is the material parameter,  $\bar{n}$ is a rate-sensitivity parameter.}  The resolved shear stress $\tau^r$ is the projection of the Mandel stress tensor $\mathbf{M}_e$ on the direction and plane of slip:
	\begin{equation}
	\tau^r=\mathbf{m}^r_0\cdot\mathbf{M}_e\cdot\mathbf{n}^r_0,\hspace{0.2cm}
	\mathbf{M}_e = \mathbf{F}^T_e \mathbf{S} \mathbf{F}^T_p = \mathbf{F}^T_e \boldsymbol{\tau} \mathbf{F}^{-T}_e \,,
\end{equation}
where $\mathbf{S}$ is the first Piola-Kirchhoff stress and $\boldsymbol{\tau}$ is the Kirchhoff stress. The Mandel stress tensor is obtained using a hyper-elastic law:
\begin{equation}
	\mathbf{M}_e=2\mathbf{C}_e\frac{\partial \Psi}{\partial \mathbf{C}_e}\,,
\end{equation}
where $\mathbf{C}_e=\mathbf{F}^T_e\mathbf{F}_e$ is the right elastic Cauchy-Green tensor and
\begin{equation}
	\Psi=\frac{1}{2}\mathbf{E}_e\cdot\mathbb{L}^e\cdot\mathbf{E}_e
\end{equation}
is the  {Kirchhoff-type function of} free energy density per unit volume in the reference configuration, $\mathbb{L}^e$ is the anisotropic stiffness tensor of single crystal and $\mathbf{E}_e=\frac{1}{2}(\mathbf{C}_e-\mathbf{1})$ is the elastic Lagrangian strain tensor.

The evolution of {the} critical value of the resolved shear stress is governed by the exponential Voce law:
\begin{equation}
	\dot{{\tau}_c}^r = H(\Gamma)\sum_{s=1}^{M}h_{rs} \left| \dot{\gamma}^s \right|,
\end{equation}
{where $h_{rs}$ is the latent hardening parameter, equal 1 for self-hardening {($r=s$)}, $q_0$ for {latent} hardening {($r \neq s$)} on coplanar systems ($\mathbf{n}^r_0\cdot\mathbf{n}^q_0=1$), and $q$ for {latent} hardening on non-coplanar systems ($\mathbf{n}^r_0\cdot\mathbf{n}^q_0\neq 1$) and
	\begin{equation}
		H(\Gamma)=\frac{\rm{d}\tau_c(\Gamma)}{\rm{d}\Gamma},
		\hspace{0.2cm}
		\tau_c(\Gamma)=\tau_0+\left(\tau_1+\theta_1 \Gamma \right)
		\left(1-\exp\left({-\Gamma \frac{\theta_0}{\tau_1}}\right) \right)
	\end{equation}
	\begin{equation}
		\Gamma=\int \dot{\Gamma} {\rm{d}}t,
		\hspace{0.2cm}
		\dot{\Gamma}=\sum_{r}{|\dot{\gamma}^r|}
	\end{equation}
	
	The parameters of the hardening model, elastic constants of the material, and the value of $n$ used are shown in the table \ref{tab:param}. The latent hardening parameter on both coplanar and non-coplanar systems is the same, but in general, it could have been taken as different.}

\begin{table*}
	\centering
	\begin{tabular}{|c|c|c|c|c|c|c|c|c|c|c|}
		\hline
		$C_{11}$   & $C_{12}$   & $C_{44}$ & $\tau_0$ & $\tau_1$ & 
		$\theta_0$ & $\theta_1$ & $q$      & $q_0$    & $n$  & $v_0$\\
		GPa &  GPa   & GPa & GPa  & GPa  & GPa
		&  GPa &     &     &  &\\
		\hline
		150 & 75 & 37.5 & $0.02$ & $0.097$ & $0.18$ & ${0.0\times10^{-3}}$ & 1.4 & 1.4 & 10 & 0.001\\
		\hline
	\end{tabular}
	\caption{\; Elastic constants ($C_{11}, C_{12}, C_{44}$), initial critical shear stress ($\tau_0$), and hardening model parameters ($\tau_1, \theta_0, \theta_1, q, q_0$), exponent in the power law (n) and reference shear rate ($v_0$) \citep{POTIRNICHE2006921}. }
	\label{tab:param}
\end{table*}

\subsection{FE implementation}\label{2.2}
The standard procedures developed for the FE implementation of finite strain elasto-plasticity in the fully Lagrangian displacement-based setting are followed \citep{Simo98}. In particular, incremental constitutive equations have been obtained by applying the implicit  backward-Euler  time  integration  scheme and the relation (\ref{Eq:Fpevolution}) is integrated using the exponential map, 
\begin{equation}
	\mathbf{F}_p(t+\Delta t)=\exp(\Delta t\hat{\mathbf{L}}_p)\mathbf{F}_p(t).
\end{equation}
{The implementation has been performed using AceGen code generator \citep{Korelc02}. It combines the symbolic algebra capabilities of Wolfram Mathematica with automatic differentiation and advanced techniques of expression optimization. The package enables  straightforward derivation of an algorithmic consistent tangent that leads to a quadratic convergence rate. Computations were performed using the AceFEM package. In calculations, 4-noded linear quadrilateral elements with 4 integration points are used. Additionally, the F-bar method \citep{SouzaNeto96} is applied in order to have a robust implementation, enabling the enforcement of nearly incompressible material behaviour in the geometrically non-linear regime. In spite of considering the 2D plane strain problem, the three-dimensional nature of the crystal plasticity model, and specifically the geometry of slip systems, are fully taken into account. At each Gauss point the material displacement gradient $\mathbf{H}=\mathbf{F}-\mathbf{I}$ is assumed for which components $H_{i3}=H_{3i}=0$ ($i=1,2,3$) and '3' denotes the direction perpendicular to the plane.} 

\section{Unit cell model and boundary conditions}
\label{Unitcellmodelandbc}

\subsection{Cell model}

\begin{figure*}[!t]
    \centering
    	\begin{subfigure}[!htp]{0.496\textwidth}
		\centering
		\includegraphics[angle=0,width=\textwidth]{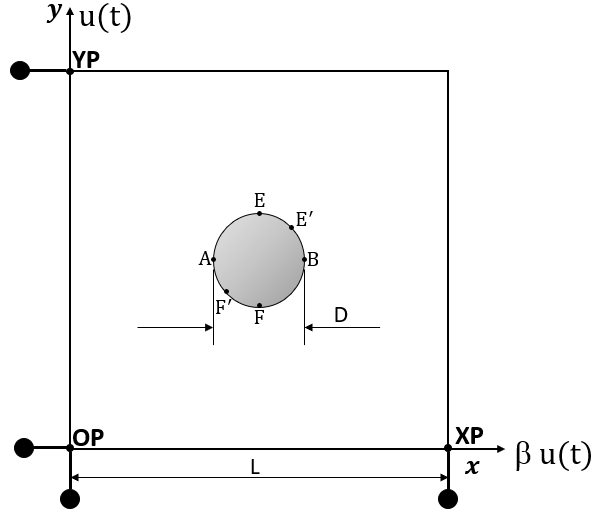}
		\caption{}
		\label{schematicdisplacementcontrolled}
	\end{subfigure}
	\begin{subfigure}[!htp]{0.496\textwidth}
		\centering
		\includegraphics[angle=0,width=\textwidth]{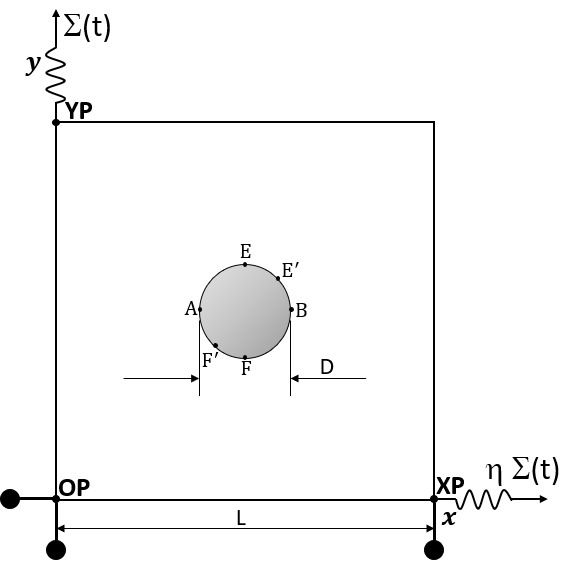}
        \caption{}
		\label{schematicstresscontrolled.PNG}
	\end{subfigure}
	\caption{\;Schematic representation of model (a) Displacement controlled (b) Stress controlled. D/L= 0.2 is imposed in all analyses corresponding to 0.0314 void volume fraction.}
    \label{unitcellmodel}
\end{figure*}

A 2D plane strain unit cell with one cylindrical void is employed. Cartesian coordinate system ($x-y$) is used and the origin of the coordinate frame is placed at the node {OP}. The initial diameter of the void is $D$ and the square plate has a side length of $L$. The ratio of $D/L$ is used to define the void volume fraction: $f=\pi/4(D/L)^2$. Nodes OP, XP, and YP are used to prescribe the boundary conditions. Instead of confining the sides of the unit cell to stay planar, which can over-constrain the model leading to the development of high stresses for some orientations, the periodic boundary conditions are applied. Accordingly, the displacement of corresponding nodes on opposite sides of the unit cell in the $x-y$ plane are connected by periodic boundary conditions, namely

\begin{equation}\label{Eq:periodicbc}
	\mathbf{u_2 - u_1 = \bar{H} (x_2 - x_1)},
\end{equation}
where $\bar{\mathbf{H}}$ is the overall (averaged) material displacement gradient tensor of the unit cell, $\mathbf{u_1, u_2}$ represent the displacement of corresponding nodes on opposite sides and $\mathbf{x_1, x_2}$ represent the corresponding nodal vectors at the reference configuration.

Several studies were carried out by \cite{HAN20132115,SRIVASTAVA20131169,SRIVASTAVA201510,KOPLIK1988835} on unit cells containing voids imposed with constant stress triaxiality ratio (ratio of the mean stress to the von Mises stress) boundary conditions. On the other hand, strain-controlled boundary conditions were considered by
 \cite{SCHACHT20031605,POTIRNICHE2006921}. In the present study for overall loading, both kinds of boundary conditions are employed in order to compare and quantify the influence of strain and stress biaxiality ratio on the void growth and coalescence. The way in which they are imposed is described in the next subsections.

\subsection{Displacement controlled boundary conditions}

For the displacement controlled boundary conditions, a displacement biaxiality factor $\beta$ is set, which is defined as the ratio of the displacement in the  $x$ direction to the displacement in the $y$ direction, namely $\beta=u_x(\rm{XP})/u_y(\rm{YP})=const$. Therefore, the following displacement boundary conditions are imposed at the reference configuration as shown in Figure \ref{schematicdisplacementcontrolled}:
\begin{itemize}

\item at node OP, $u_x = u_y = 0$,

\item at node XP, $u_x = \beta u(t), u_y = 0$,

\item at node YP, $u_x = 0, u_y = u(t) $,

\end{itemize}
which result in the following components of the displacement gradient $\bar{\mathbf{H}}$ in Eq. (\ref{Eq:periodicbc}) 
\begin{displaymath}
{\bar{H}}_{kl}=\frac{u(t)}{L}\left[\begin{array}{ccc}
\beta & 0 & 0 \\
0 & 1 & 0\\
0 & 0 & 0\\
\end{array}\right]
\end{displaymath}
Note that all components of $\bar{\mathbf{H}}$ are known for this loading scenario.

For the uniaxial tension/compression case in the $y$-direction of the sample, considered at the beginning of the next section, the following displacement boundary conditions are imposed: 
\begin{itemize}
\item at node OP, $u_x = u_y = 0$
\item at node XP, $u_y = 0$
\item at node YP, $u_y = u(t) $,
\end{itemize}
which result in the following components of the displacement gradient $\bar{\mathbf{H}}$ in Eq. (\ref{Eq:periodicbc})
\begin{equation}\label{Eq:uniaxial-disp}
{\bar{H}}_{kl}=\frac{u(t)}{L}\left[\begin{array}{ccc}
\star & \star & 0 \\
0 & 1 & 0\\
0 & 0 & 0\\
\end{array}\right]\,,
\end{equation}
whereby $\star$ we denoted unknown components of $\bar{\mathbf{H}}$. By energy minimization, this leads to the averaged Cauchy stress for which $\Sigma_{xy}=\Sigma_{xx}=0$, so the stress biaxiality factor $\eta=\Sigma_{xx}/\Sigma_{yy}=0$. Note that this case is not equivalent to the 3D uniaxial tension case since, in general all $\Sigma_{kz}$, $k=x,y,z$ are not necessarily zero for anisotropic material under plane strain conditions. 

\begin{figure*}[t!]
    \centering
    \includegraphics[angle=0,width=0.5\textwidth]{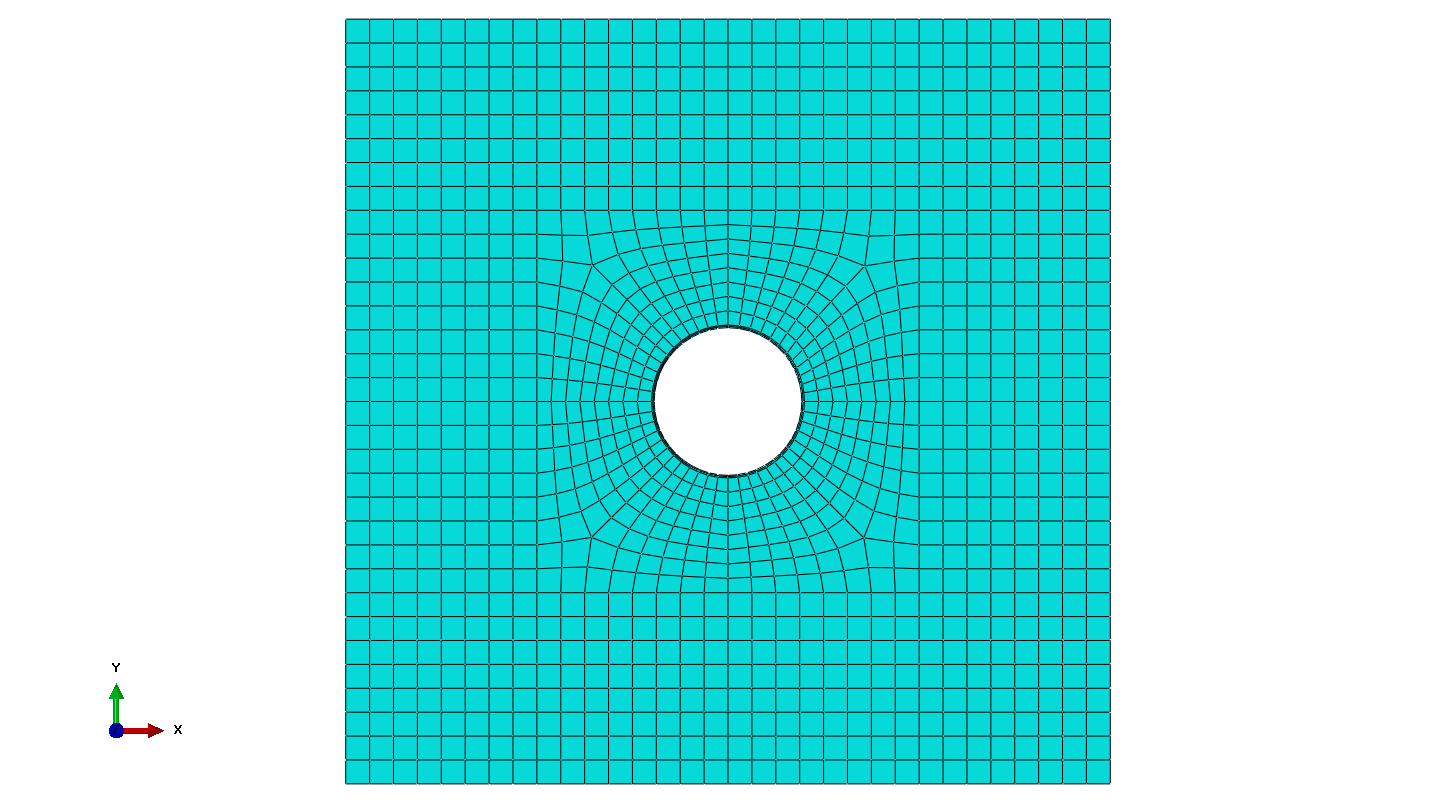}
    \caption{Finite element mesh with a cylindrical void.}
	\label{fig:mesh}
\end{figure*}

\subsection{In-plane stress controlled boundary conditions}

For controlling the in-plane stress biaxiality factor $\eta$, the formulation based on the proposal by \cite{LING201658} is employed in the present study.  A special spring element oriented in the direction of principal loading is employed to regulate displacement at the nodes XP and YP in order to maintain the constant stress ratio. Node OP is fixed and the displacement of node XP in the $y$ direction is disabled to remove rigid motion as shown in the figure \ref{schematicstresscontrolled.PNG}. In-plane stress biaxiality $\eta$, which is defined as the ratio of the Cauchy's stress normal components along $x$ direction to $y$ direction, namely $\eta = \Sigma_{xx}/\Sigma_{yy}=\rm{const}$ is kept constant to study the void growth.  Application of the element results in the averaged Cauchy stress for the unit cell of the form,
\begin{displaymath}
{\Sigma}_{kl}=\Sigma_{yy}(t)\left[\begin{array}{ccc}
\eta & 0 & \star \\
 & 1 & \star\\
\rm{sym.} &  & \star\\
\end{array}\right]\,,
\end{displaymath}
whereby $\star$ we denote unknown components of $\boldsymbol{\Sigma}$. Note that $\Sigma_{yy}(t)$ is also unknown, while via the truss element, displacement $u_y(\rm{YP})$ is imposed.

Different displacement and in-plane stress biaxiality ratios: $\beta$ and $\eta$ are considered in the present study, which are compared and discussed in the next section.

\subsection{Finite element geometry and mesh}

Two commercial software packages are used in this study. 2D planar model and mesh are generated using the commercial CAE software (ABAQUS version 6.13) as      shown in Fig.~\ref{fig:mesh}. Then the mesh data is imported to a symbolic and algebraic system, Wolfram Mathematica as specified in section \ref{2.2} to perform finite element calculations and post-processing using the AceFEM package. The ratio of the void diameter to the side length in the x-y plane is taken as  D/L = 0.2, leading to an initial void volume fraction of $f = 0.0314$. 
2D mesh is employed with 1168 elements of type CPE4R. Mesh convergence tests were carried out on a number of unit cells with different mesh sizes. Convergence was evaluated by determining the evolution of the relative void volume fraction with the overall effective strain. 

Similarly to other studies (see e.g. \citep{LING201658}), the overall Cauchy stress $\boldsymbol{\Sigma}=\frac{1}{\bar{J}}\bar{\mathbf{S}}\bar{\mathbf{F}}$ ($\bar{J}=\det\bar{\mathbf{F}}$) is calculated based on the volume averaged first Piola-Kirchhoff stress $\bar{\mathbf{S}}$  found as:
\begin{equation}
    \bar{\mathbf{S}}=\frac{1}{V}\int_V\mathbf{S}(\mathbf{X})dV=\frac{1-f}{V_m}\int_V\mathbf{S}(\mathbf{X})dV_m\,,\quad
\end{equation}
where $V_m$ is the bulk crystal volume and the integration is performed numerically in the reference configuration. Unknown components of the deformation gradient $\bar{\mathbf{F}}=\mathbf{I}+\bar{\mathbf{H}}$ are calculated based on the relation (\ref{Eq:periodicbc}) using the current displacement vectors at nodes XP and YP, namely
\begin{equation}
\bar{F}_{11}=1+\frac{u_x({\rm{XP}})}{L}\,,\quad \bar{F}_{21}= \frac{u_y({\rm{XP}})}{L}\,,\quad \bar{F}_{12}= \frac{u_x({\rm{YP}})}{L}\,,\quad \bar{F}_{22}=1+\frac{u_y({\rm{YP}})}{L}\,.    
\end{equation}
%%%%%%%%%
\begin{table}[t!]
    \centering
    \begin{tabular}{|l|c|c|c|}
    \hline
    \multirow{2}{4em}{Crystal Orientation} & Lateral& Primary & Lateral \\
 & direction ($x$)& loading direction ($y$)&direction ($z$)\\
    \hline
    Orientation O&[$\bar{1}$10]&[001]&[110]\\
    Orientation A&[111]&[$\bar{2}$11]&[0$\bar{1}$1]\\
    Orientation B &[100]&[010]&[001]\\
     Orientation C &[110] &[$\bar{1}$10]&[001]\\
    \hline
    \end{tabular}
    \caption{\; Crystal orientations considered with respective global coordinate axes.}
    \label{tab:crystal orientations}
\end{table}
%% Results and discussion
\section{Results and Discussion}

In this section, on the basis of the results of finite element simulations, the impacts of crystallographic orientation and various boundary conditions on the void growth or coalescence as well as grain refinement due to heterogeneous lattice rotation, in a 2D plane strain unit cell are examined. {First, in-plane uni-axial compression and tension, simulated using the displacement controlled scenario (Eq. \ref{Eq:uniaxial-disp}), are performed to demonstrate the void-induced heterogeneous slip activity, which then leads to spatial variation in lattice rotation. The example serves also to explore the effect of loading direction with respect to crystal axes and loading 'sign' (tension vs. compression).  Moreover, the analysis preliminary verifies the model predictions with available experimental findings provided in \citep{Gan06}.} Next, various in-plane biaxial processes are considered. The displacement-controlled boundary condition will be referred to as the $\beta$ loading case and the stress-controlled boundary condition will be referred to as the $\eta$ loading case throughout the discussion. To see how plastic anisotropy impacts void growth in an FCC single crystal, four initial orientations of the crystalline lattice with respect to the sample axes are taken into consideration. They are collected in Table \ref{tab:crystal orientations}.  In the current study, seven loading scenarios with $\beta$ equal to -0.5, 0, and 1 as well as $\eta$ equal to -0.5, 0, 0.8, and 1 are analysed. The scenario $\eta=0.8$ is selected due to its approximate equivalence to the $\beta=0$ case. Note that the state of in-plane uni-axial tension or compression is represented by $\eta$ = 0. 

\subsection{Microstructure evolution and void growth in in-plane uni-axial tension and compression}

\cite{Gan06} examined in-plane uni-axial compression of a single crystal along the [001] direction with a cylindrical void axis along the [110] (orientation O in Table \ref{tab:crystal orientations}). As discussed in detail by \cite{Gan06}, by applying the anisotropic rigid-plastic slip line theory, this configuration ensures a plane strain condition in the [001] - [$\bar{1}$10] crystal plane under the action of the  compressive or tensile loading with three effective in-plane slip systems. 
{For the pristine crystal they are results of equal activity of four systems (see Table \ref{tab:active-systems}), which act in opposite directions (i.e. $\mathbf{m}$ and $-\mathbf{m}$) for tensile and compressive loading in the plane. It could be also verified that when the direction of loading is changed to [001], under a plane strain condition, the same set of slip systems will be active, again in the opposite sense. Thus, as far as plastic deformation by dislocation motion is considered, in-plane compression (cor. tension) in [001] is equivalent to in-plane tension (cor. compression) in [$\bar{1}$10].}

\begin{table}[t!]
    \centering
    \begin{tabular}{|l|cccc|}
    \hline
    \multirow{2}{4em}{Loading} & \multicolumn{4}{|c|}{Slip system} \\ 
& $(1\bar{1}1)[10\bar{1}]$& $(1\bar{1}1)[011]$ & $(\bar{1}11)[101]$ &$(\bar{1}11)[01\bar{1}]$ \\
    \hline
    Tension [001] & $-\gamma$ & $\gamma$ & $\gamma$ & $-\gamma$ \\
    Compression [001] & $\gamma$ & $-\gamma$ & $-\gamma$ & $\gamma$\\
    Tension [$\bar{1}$10] & $\gamma$ & $-\gamma$ & $-\gamma$ & $\gamma$\\
    Compression [$\bar{1}$10] & $-\gamma$ & $\gamma$ & $\gamma$ & $-\gamma$ \\
    \hline
    \end{tabular}
    \caption{\;Set of active slip systems for the specified in-plane uniaxial loading of Orientation~O (no void) according to the rigid-plastic crystal plasticity model. $\pm\gamma$ denotes the magnitude of the slip on the specified system and its sign at the same level of the true strain in the given loading direction.}
    \label{tab:active-systems}
\end{table}

For the sample with a cylindrical void, under the same loading conditions, the formation of regions of unequal slip activity of potentially active systems around the void is observed, which leads to the lattice rotation heterogeneity and crystal fragmentation into subgrains. These theoretical predictions were verified by \cite{Gan06} experimentally, for the [001] compression case, by EBSD measurements. Note that for the crystal without the void, no lattice rotation is predicted by the model, and slip activity is homogeneous, so the grain is not fragmented.

{The purpose of the study is to investigate the differences between void evolution and grain fragmentation using four loading scenarios, namely tension/compression in the [$\bar{1}$10] and [001] directions, even though the same active slip systems are expected for all cases in a pristine rigid-plastic crystal (as indicated in Table \ref{tab:active-systems}). 
It is obvious that the overall stress biaxiality factor $\eta$ is equal to 0 in each case. }

%% figures
\begin{figure*}[t!]
    \centering
    \includegraphics[angle=0,width=\textwidth]{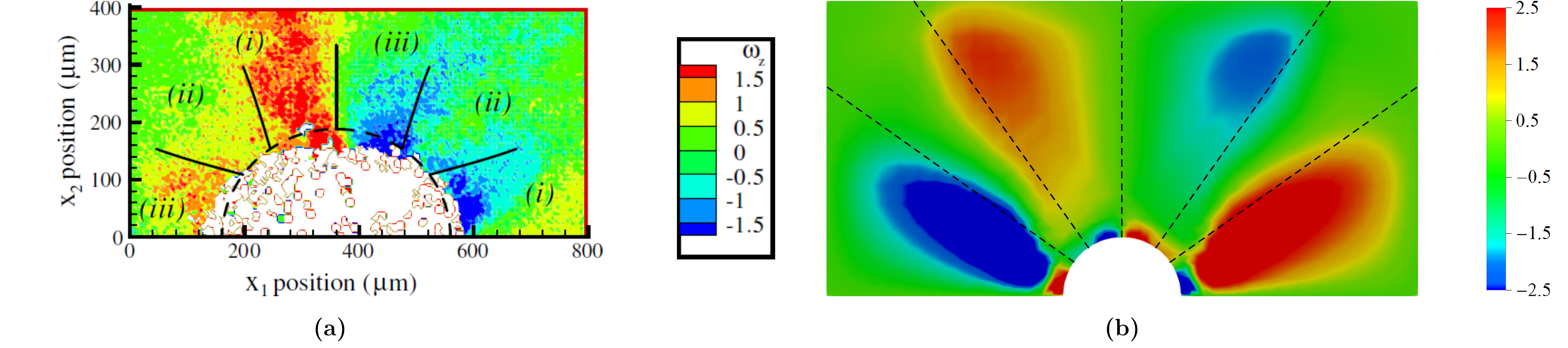}
    \caption{a) In-plane lattice rotation angle obtained using EBSD measurement (reprinted from \citep{Gan06} with permission from Elsevier) and b) misorientation angle map plotted found by the CPFEM method for a 5\% compression strain. The dashed line represents the slip sectors at an angle of 35.3, 54.7, and 90 degrees respectively.}
    \label{fig:ganetal-rotangsector.pdf}
\end{figure*}

First, we have used this example to verify the predictive capabilities of the present numerical model. As shown in \cite{Gan06} and confirmed in our study, one of three effective in-plane slip systems dominates in three different angular slip sectors which are centred at the middle of the void, as marked by dashed lines in Fig. \ref{fig:ganetal-rotangsector.pdf}b. This results in different lattice rotations in respective domains. In figure \ref{fig:ganetal-rotangsector.pdf}b, we present misorientation angle distribution\footnote{See definition (\ref{Eq:miorient-angle}), which is in the present case equipped with the sign to indicate the in-plane rotation direction. Sign $+$ denotes clockwise and $-$ anticlockwise rotation of [001] axis} for the compression strain of 5\%.  Qualitatively similar subdivision is seen in experimental data quoted in figure \ref{fig:ganetal-rotangsector.pdf} after \cite{Gan06}. There are some differences concerning the direction of rotation in the lateral domains, however, a full quantitative comparison is not possible due to the lack of the detailed experiment geometry and boundary condition data in \citep{Gan06}. 

\begin{figure*}[t!]
    \centering
    \begin{tabular}{ccc}
   a)\includegraphics[angle=0,width=0.3\textwidth]{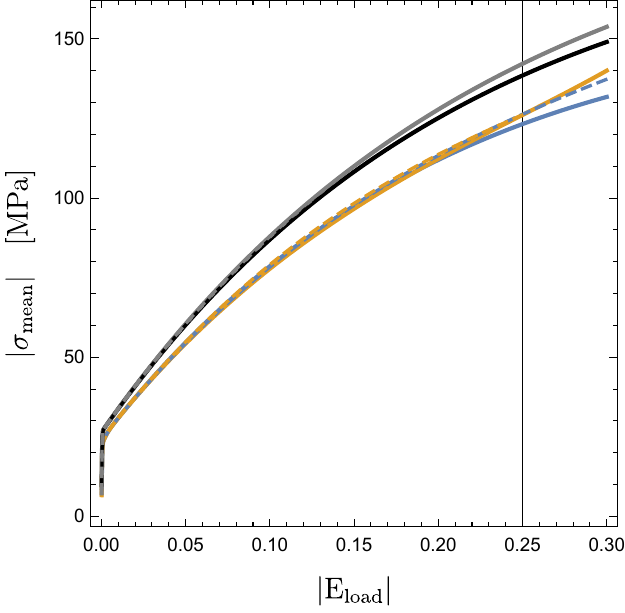}&
    b)\includegraphics[angle=0,width=0.3\textwidth]{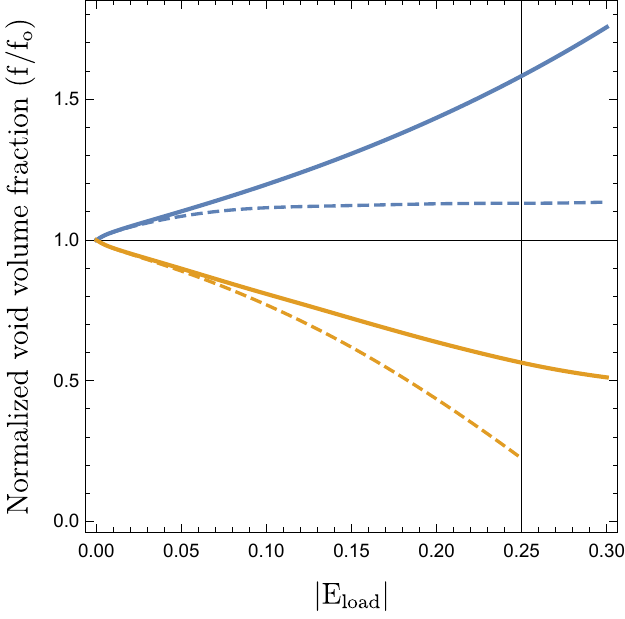}&\includegraphics[angle=0,width=0.25\textwidth]{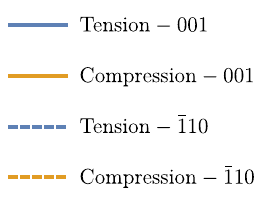}\\
    c)\includegraphics[angle=0,width=0.3\textwidth]{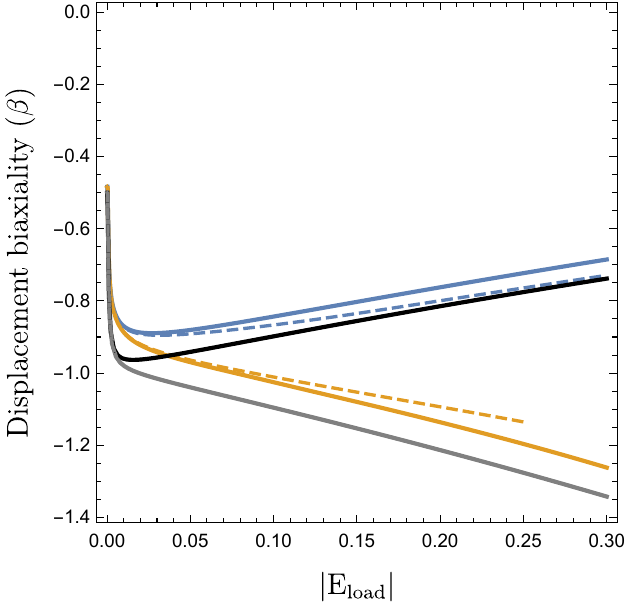}&
    d)\includegraphics[angle=0,width=0.3\textwidth]{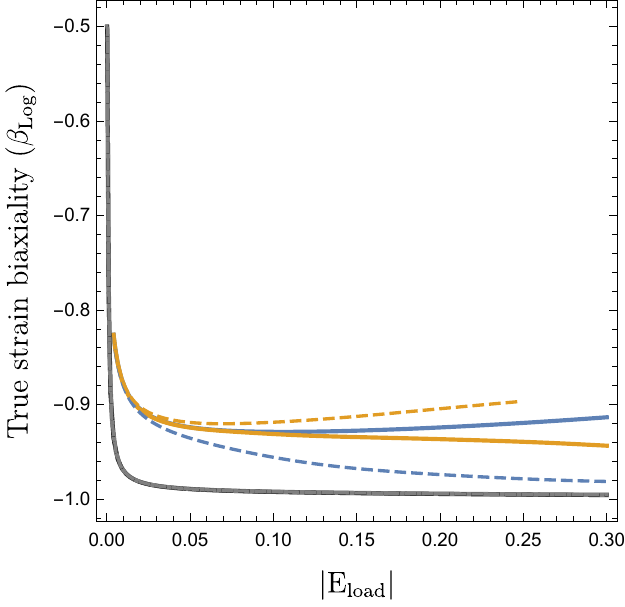}&\includegraphics[angle=0,width=0.3\textwidth]{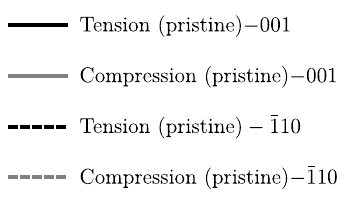}\\
    \end{tabular}
    \caption{Comparison of a) absolute value of in-plane mean stress ($\sigma_{\rm{mean}}$)b) normalized void volume fraction c) displacement biaxiality factor ($\beta$), d) true strain biaxiality factor ($\beta_{\rm{log}}$) for four loading cases between pristine and voided single crystal. Evolution of quantities is presented as a function of the absolute value of, $E_{\rm{load}}=\ln(1+u/L)$ where $u$ is the displacement in the loading direction.}
    \label{fig:averload-fourcases}
\end{figure*}

\begin{figure*}[t!]
    \centering
    \begin{tabular}{ccc}
    a)\includegraphics[angle=0,width=0.4\textwidth]{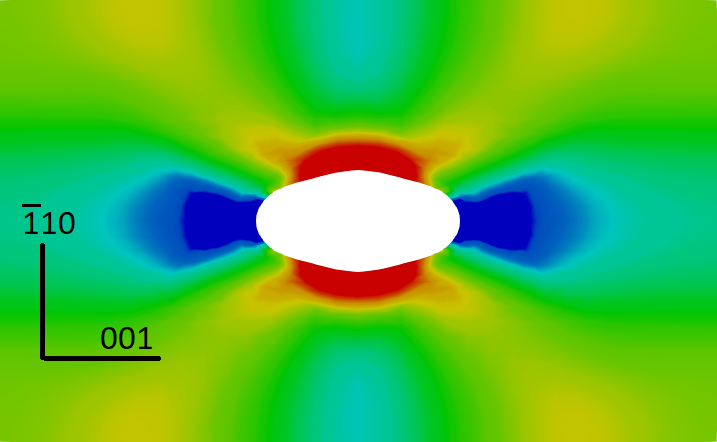}&
    b)\includegraphics[angle=0,width=0.4\textwidth]{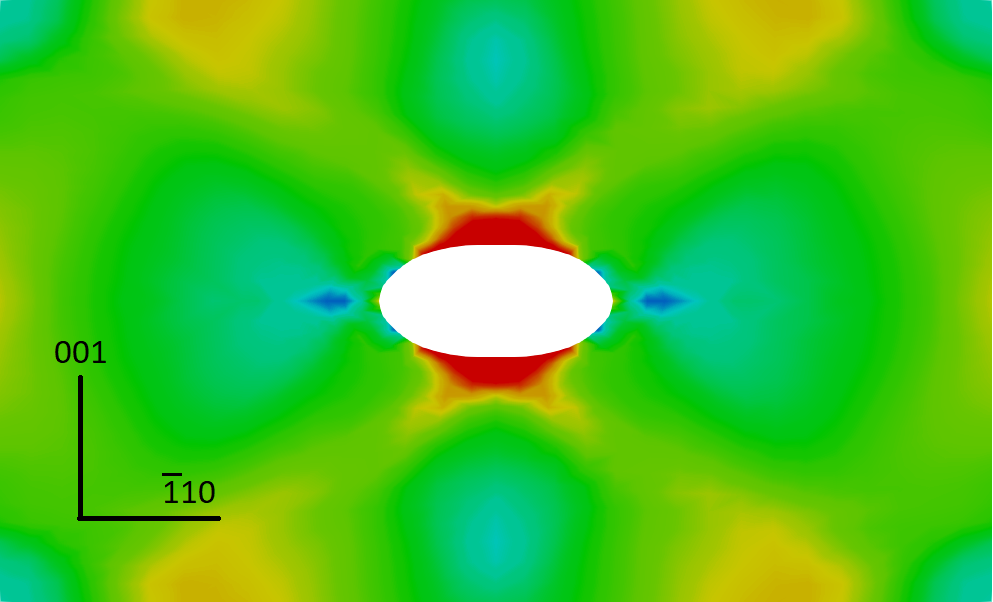}&\\
    c)\includegraphics[angle=0,width=0.4\textwidth]{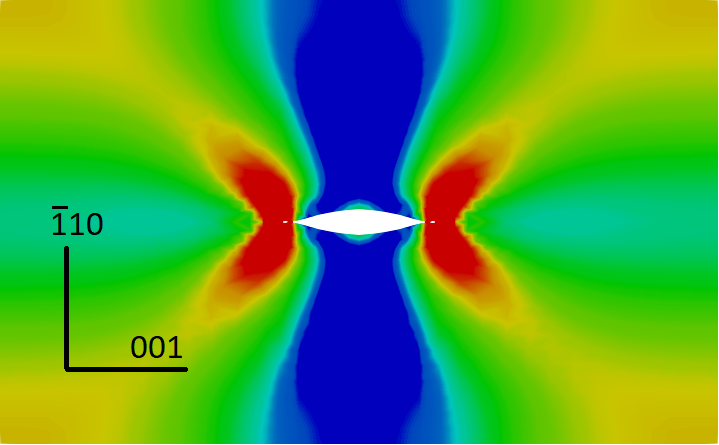}&
     d)\includegraphics[angle=0,width=0.4\textwidth]{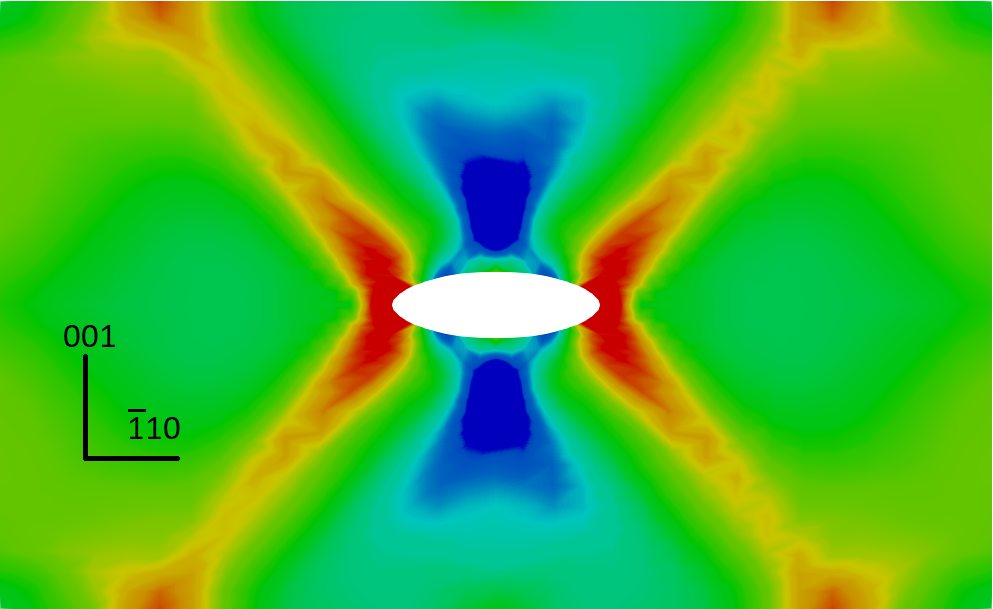}&\includegraphics[angle=0,width=0.05\textwidth]{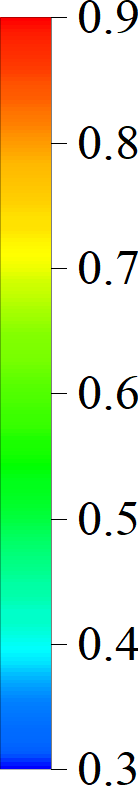}\\
    \end{tabular}
    \caption{Space distribution of accumulated shear at true strain $|E_{\rm{load}}|$ level 0.25. a) Tension[001] b) Tension [$\bar{1}$10] c) Compression [$\bar{1}$10], d) Compression [001]. Initial orientation of crystallographic directions [001] and [$\bar{1}$10] was marked on the plots.}
    \label{fig:accumshear-fourcases}
\end{figure*}
\begin{figure*}[t!]
    \centering
    \begin{tabular}{ccc}
   a) \includegraphics[angle=0,width=0.4\textwidth]{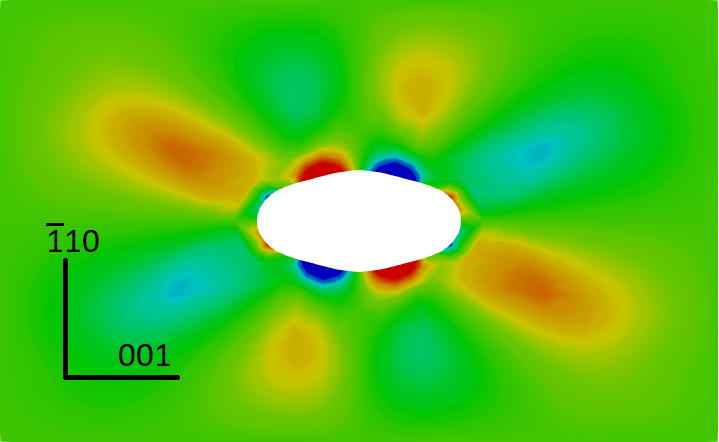}&
    b)\includegraphics[angle=0,width=0.4\textwidth]{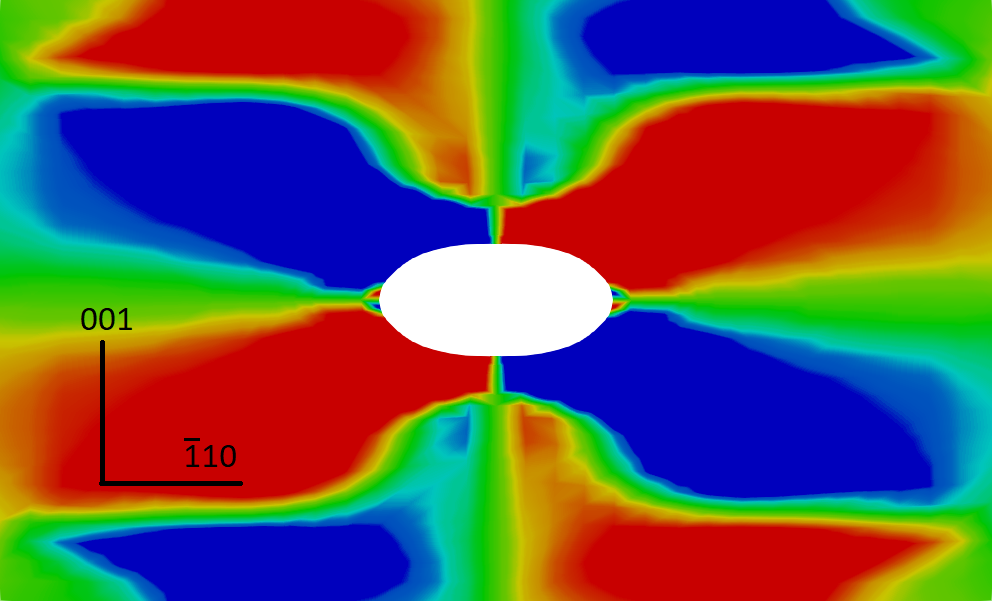}&\\
    c)\includegraphics[angle=0,width=0.4\textwidth]{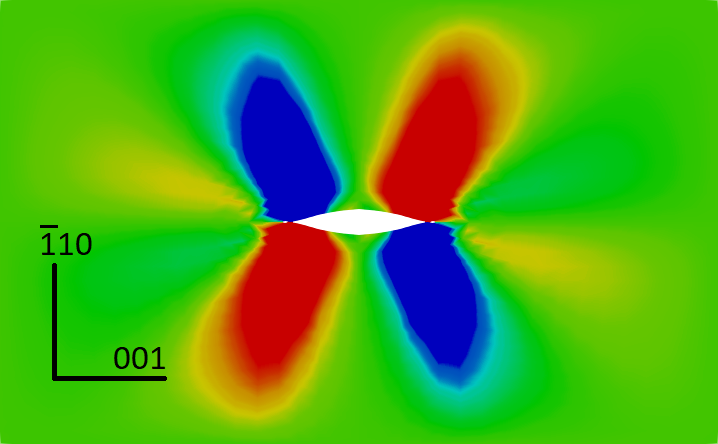}&
    d)\includegraphics[angle=0,width=0.4\textwidth]{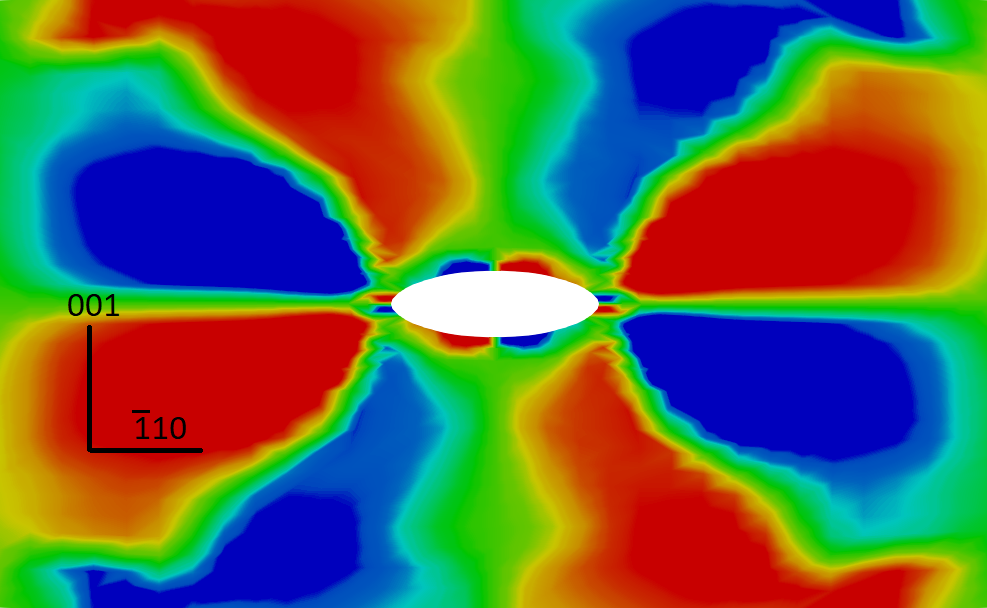}&\includegraphics[angle=0,width=0.05\textwidth]{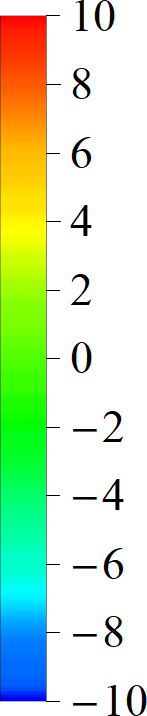}\\
    \end{tabular}
    \caption{Local lattice rotation angle at true strain $|E_{load}|$ level of 0.25 in the loading direction. a) Tension [001] b) Tension [$\bar{1}$10] c) Compression [$\bar{1}$10], d) Compression [100]. Initial orientation of crystallographic directions [001] and [$\bar{1}$10] was marked on the plots.}
    \label{fig:rotangle-fourcases}
\end{figure*}

{Next, the same sample configuration is used to explore the effect of loading direction and its sign (i.e. in plane tension vs. compression) on the void evolution and grain refinement}. To this end, four processes enlisted in Table \ref{tab:active-systems} are studied numerically.
In figure \ref{fig:averload-fourcases}a, we compare the overall in-plane mean stress variation vs. magnitude of the true strain in the loading direction. It is seen that initially, the response in terms of the magnitude of the in-plane mean stress ($\sigma_{\rm{mean}}=1/2(\Sigma_{xx}+\Sigma_{yy})$) is the same for all processes and does not show visible tension-compression asymmetry. However, as the deformation proceeds, the difference starts to increase due to differences in the lattice rotation and void evolution. In each case, the stress level is smaller for the porous crystal than for the pristine one. The evolution of the normalized void volume fraction ($f/f_0$) is presented in figure \ref{fig:averload-fourcases}b. It is observed that, as expected, the void volume fraction increases for tension and decreases for compression, however, there are important differences between the two loading directions. While for tension in [001] void grows monotonically, for tension in [$\bar{1}$10] after initial increase void volume stabilizes at some, relatively small, constant value ($f/f_0\sim 1.13$), at least for the demonstrated strain regime\footnote{It has been verified that for both these loading cases, the void volume starts to increase with accelerating rate and softening is observed for in-plane mean stress at larger strain,  which events eventually lead to void coalescence, see Fig. S.1 in the supplementary material\label{foot:col}.}. On the other hand, for compression in [$\bar{1}$10] void volume decreases monotonically and void starts to close at a relatively small true strain level ($\sim{0.25}$)\footnote{Calculations were stopped at the moment when the void opposite boundaries were first in contact since the material overlapping was not prevented in calculations.}, and for compression in [001] after initial important decrease the void collapse is postponed to higher strain values. It should be stressed that the overall stress triaxiality value, calculated accounting for the 3D character of the stress field (note that $\Sigma_{zz}$ is not zero for all analysed cases), is approximately equal to 0.5 for both tension loadings and -0.5 for compression. Only small variations in the Lode parameter calculated for the overall stress are detected for four loading cases (its value is around 0.32-0.33). 

Displacement biaxiality ratio $\beta$, calculated here as the inverse ratio of in-plane displacements in loading direction with respect to the lateral one, is seen in Fig. \ref{fig:averload-fourcases}c, while the in-plane true strain biaxiality $\beta_{\rm{log}}$, calculated as the corresponding ratio of in-plane components of true strain measure (e.g. for tension/compression in [001] it is $\beta_{\rm{log}}=E_{xx}/E_{yy}=\ln(F_{xx})/\ln(F_{yy})$) is shown in Fig. \ref{fig:averload-fourcases}d. Their variation with strain is compared for all four processes and pristine and voided crystals. As expected, it is observed that for a crystal without a void for all processes the evolution of $\beta_{\rm{log}}$ is the same: it starts with the value of -0.5 in the elastic regime and reaches -1.0 for well-developed plastic flow, which marks incompressible deformation in that regime. On the other hand, for voided crystals, the value of -1 is approached only for tension in [$\bar{1}$10], which is related to the stabilization of the void growth. For the remaining processes, the value does not drop below -0.95 indicating the compressibility of voided crystal. 

Differences in the void growth or closing for two loading directions concern also the developed void shape as seen in Figs \ref{fig:accumshear-fourcases} and \ref{fig:rotangle-fourcases}. While the ellipsoidal shape of the void is observed for tension in [$\bar{1}$10] and compression in [001] (equivalent in terms of slip activity pattern in pristine crystal), the polygonal shapes are the results of compression in [$\bar{1}$10] and tension in [001] directions. Accumulated shear maps also show the failure mode for each case. In compression cases the failure proceeds by accumulated shear localization in two intersecting bands. For tension, although at the initial stage two bands are also visible, the void coalescence takes place, much later for $[\bar{1}10]$ than for $[100]$ case (see footnote \ref{foot:col} and Fig. S.1 in Supplementary Material).

Figure \ref{fig:rotangle-fourcases} shows an interesting interplay between the void evolution and the grain fragmentation phenomenon. It is seen that for the two cases for which the void growth/collapse is halted or retarded (tension in [$\bar{1}$10] and compression in [001], respectively) the clear checker-board-type subdivision of initial grain into subgrains, misoriented with respect to each other by the angle as large as $\sim 20^o$ at the true strain level $0.25$, is found. On the other hand, for two other processes, the significant lattice rotation is seen only in the domains of intensive strain. These latter results confirm microstructure evolution as an important effect accompanying the deformation of voided crystalline materials.

{The analysis showcased in this section illustrates that both in-plane stress biaxiality and stress triaxiality, as well as displacement or strain biaxiality, alone are inadequate in determining the growth of voids. This is particularly true when anisotropic materials are analysed. It is important to note that microstructure evolution plays a substantial role in this process. Fragmentation of bulk crystal surrounding the void into subgrains may lead to significant impediment of the void volume changes.}

\subsection{Void growth and microstructure evolution in in-plane biaxial loading processes}

In this subsection, to further explore factors differentiating the void growth and accompanying grain fragmentation in FCC crystals, in-plane biaxial processes are considered, for three orientations A, B, and C defined in Table \ref{tab:crystal orientations}. Orientations were selected following \cite{POTIRNICHE2006921}. In order to investigate and differentiate the effect of stress and strain biaxiality seven loading scenarios with $\beta$ equal to -0.5, 0, and 1 as well as $\eta$ equal to -0.5, 0, 0.8, and 1 are analysed. Let us remark that orientation A, contrary to B and C, is non-symmetric with respect to the loading axes, thus shear strain component $E_{xy}$ (cor. shear stress component $\Sigma_{xy}$) may be observed  for $\eta$ (cor. $\beta$) loading cases even for a pristine crystal sample.

\begin{figure*}[p]
    \centering
    a)\includegraphics[angle=0,width=0.95\textwidth]{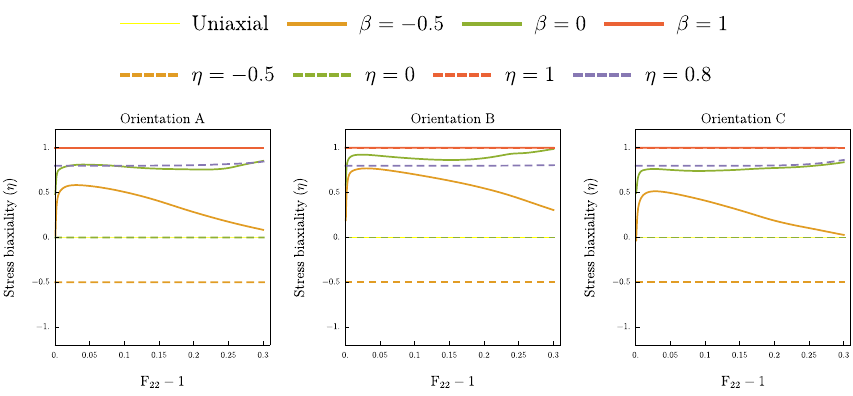}\\
    b)\includegraphics[angle=0,width=0.95\textwidth]{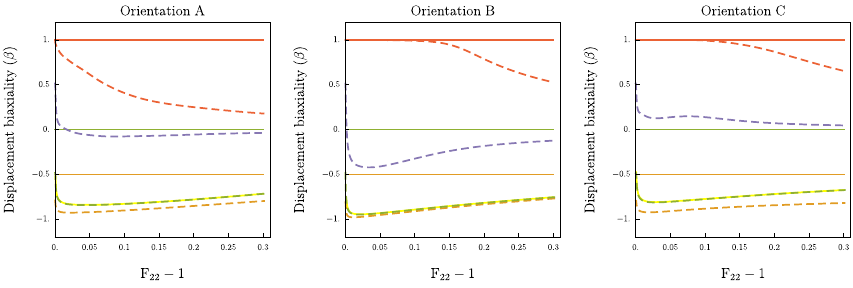}\\
    c)\includegraphics[angle=0,width=0.95\textwidth]{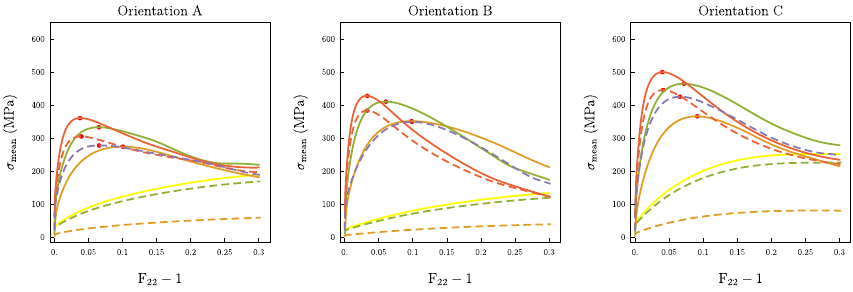}\\
    d)\includegraphics[angle=0,width=0.94\textwidth]{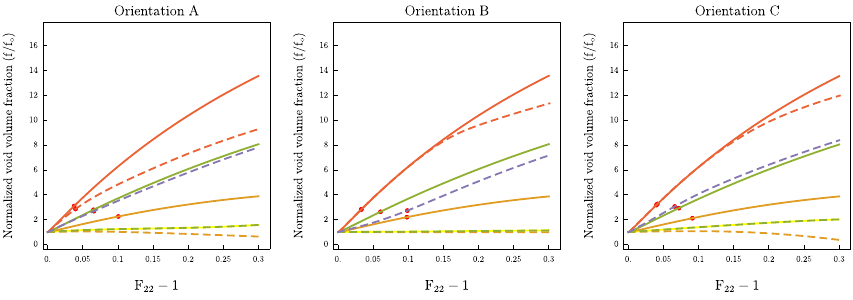}
    \caption{Variation of a) stress biaxiality ratio, b) displacement biaxiality ratio, c) overall mean stress ($1/2\; (\Sigma_{xx}+\Sigma_{yy})$) ( The peak stress is indicated by \itwas{$\bullet$}), and d) normalized void volume fraction under various loading cases.}
    \label{fig:inplanestressratioplot.png}
\end{figure*}

\begin{figure*}[t!]
    \centering
    \includegraphics[angle=0,width=\textwidth]{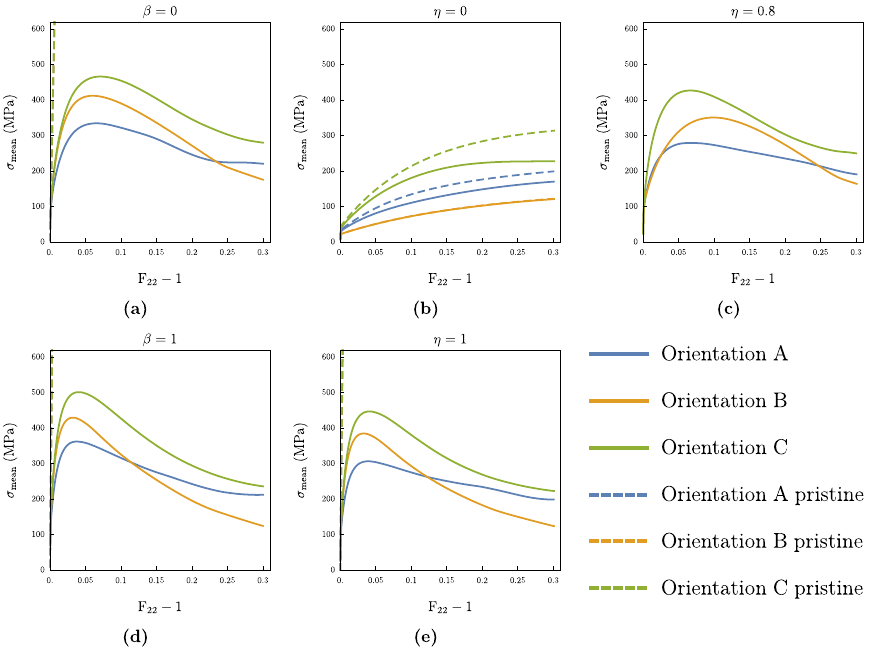}
    \caption{Overall mean stress response ($\sigma_{\rm{mean}}=1/2\; (\Sigma_{xx}+\Sigma_{yy})$) for different crystal orientations and for the loading case: a) $\beta = 0$, (b) $\eta = 0$, (c) $\eta = 0.8$  (d) $\beta = 1$ (e) $\eta = 1$.   }
    \label{fig:inplanemeanstressori.png}
\end{figure*}

\begin{figure*}[t!]
    \centering
    \includegraphics[angle=0,width=0.9\textwidth]{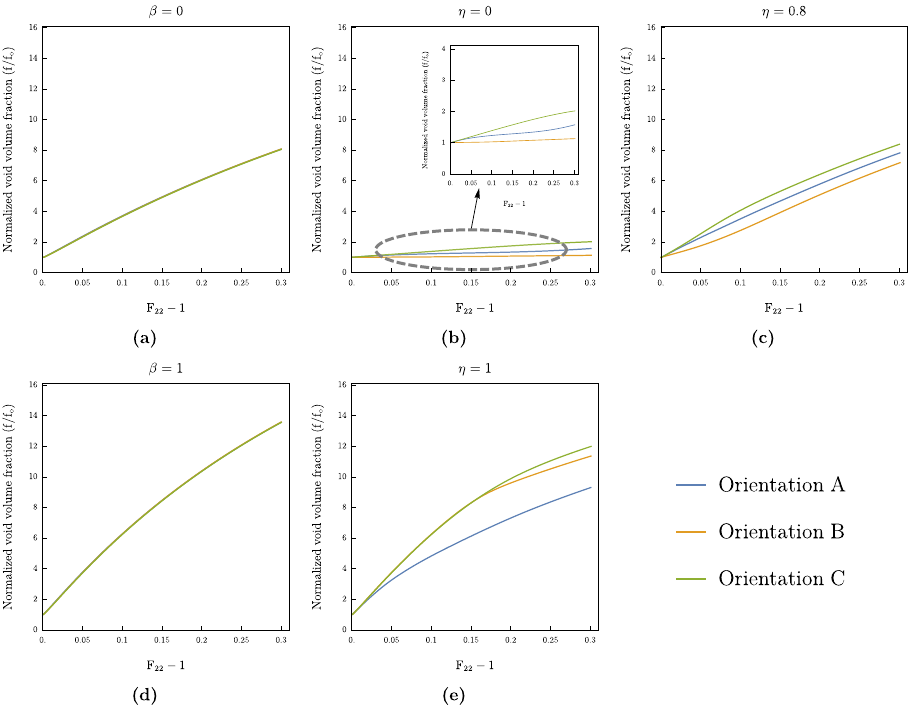}
    \caption{Normalized void volume fraction evolution for different crystal orientations and for the loading case: (a) $\beta = 0$ (b) $\eta = 0$ (c) $\eta = 0.8$  (d) $\beta = 1$ (e) $\eta = 1$. }
    \label{fig:normvoidoriplot.png}
\end{figure*}

\begin{figure*}[t!]
    \centering
      \includegraphics[angle=0,width=0.9\textwidth]{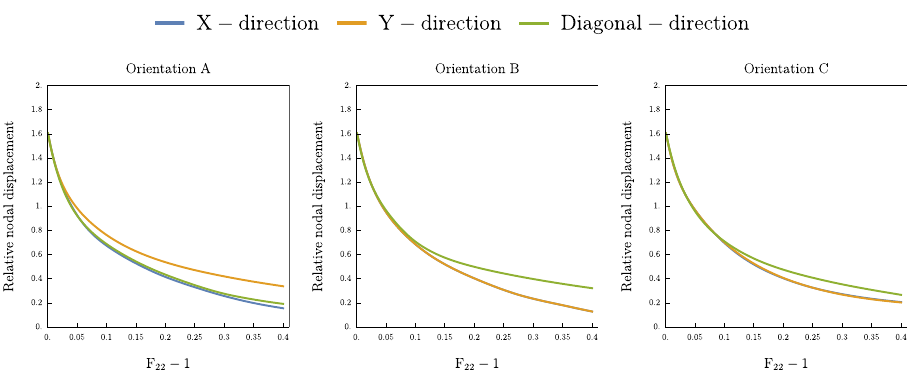}\\
      \includegraphics[angle=0,width=0.9\textwidth]{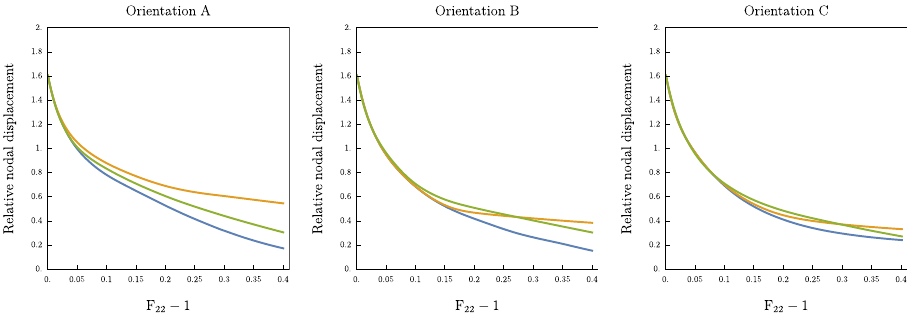}
    \caption{The change of the void diameters along X (AB), Y (EF) and diagonal (E'F') directions (see Fig. \ref{unitcellmodel}) for three orientations and two loading cases: $\beta=1$ (top) and $\eta=1$ (bottom). The curves present the value of $\log(L/|\mathbf{x}_{\rm{right}}-\mathbf{x}_{\rm{left}}|)$ where $\mathbf{x}_{\rm{right}}$ and $\mathbf{x}_{\rm{left}}$ denote current locations of nodes at the right and the left end of the respective diameter marked in Fig. \ref{unitcellmodel} while $L$ is the current cell size in a relevant direction.}
    \label{fig:coalescenceplot.png}
\end{figure*}

\subsubsection{Overall response of voided crystal}

\paragraph{Stress biaxiality ratio}
The stress biaxiality ratio for seven loading scenarios is shown in Figure~\ref{fig:inplanestressratioplot.png}a. To start with, as it is evident, the stress biaxiality ratio for the $\eta$ loading case is maintained constant for all crystal orientations during the deformation process, which verifies the validity of the finite element procedure used for imposing a constant stress biaxiality ratio. On the other hand, in general, for displacement controlled processes (with constant $\beta$) stress biaxiality ratio $\eta$ changes during the deformation process. For crystal orientations A and C, under the $\beta=-0.5$ loading case, the stress biaxiality is larger than zero; the slope initially rises, then progressively drops, and ultimately approaches the uniaxial loading case at the end of loading. Although the biaxiality ratio is marginally more than $\eta=0$ for orientation B at the end of loading, apparently, it would reach the uniaxial loading state if the deformation would have proceeded. For $\beta=0$, the slope steadily rises until it approaches $\eta=1$ at the end of loading. For this case, on average the value of stress biaxiality for three orientations is close to $\eta=0.8$ that is why for comparison purposes such stress controlled scenario is also selected for analysis. Finally, for the $\beta=1$ case,  the stress biaxiality ratio is kept constant just as it does for the $\eta=1$ case. Those graphs in conjunction with displacement biaxiality plots in Figs. \ref{fig:inplanestressratioplot.png}b are important for analysing the growth of the void and the stress response. The softening stress response is evident in Fig. \ref{fig:inplanestressratioplot.png}c when the stress biaxiality ratio increases and void growth is significant, resulting in coalescence.

Additionally, yellow lines are denoted as 'uniaxial' in Figs. \ref{fig:inplanestressratioplot.png}, show the results obtained for the in-plane uniaxial tension process without the employment of a special spring element but using the displacement-controlled conditions with $\mathbf{H}$ described by Eq. \ref{Eq:uniaxial-disp}. Calculations are performed for verification purposes and are in good agreement with the predictions obtained with the use of the spring element (marked as $\eta=0$ in figures).

\paragraph{Displacement biaxiality ratio}
The displacement biaxiality under various loading instances is depicted in Figure \ref{fig:inplanestressratioplot.png}b. Similar to the situation of stress biaxiality, the displacement biaxiality ratio $\beta$ is kept constant during the $\beta$-type process, which verifies the finite element procedure. On the contrary, in general, for stress ratio controlled processes (with constant $\eta$), the displacement biaxiality ratio varies in the course of deformation. The displacement biaxiality is kept below -0.5 for $\eta=0$ (in-plane uniaxial tension) and $\eta = -0.5$ loading cases. For asymmetric orientation A, under the $\eta=1$ loading scenario, the ratio initially follows the $\beta =1$ case, but as deformation proceeds the slope steadily falls and approaches the $\beta=0$ case. For orientations B and C, the ratio remains constant until halfway through the deformation, after which it steadily drops. For $\eta=0.8$ as expected, the strain biaxiality oscillates around $\beta=0$, although differently for each of the three orientations. For orientation A it is almost constant and close to zero, for orientation B it is negative, initially being close to -0.5 and increasing towards the uniaxial straining mode, while for orientation C it starts with a positive value and next decreases to zero. These plots are again valuable for studying in conjunction with the contour plots of accumulated shear in Section \ref{Sec:local}, void evolution (Fig. \ref{fig:inplanestressratioplot.png}d) and stress response (Fig. \ref{fig:inplanestressratioplot.png}c).

\paragraph{Overall mean stress response}
Figures \ref{fig:inplanestressratioplot.png}c illustrate the in-plane overall mean stress response for the various loading scenarios and the given crystallographic orientation. When all loading scenarios are compared, $\beta=1$ exhibits the stiffest response in the initial deformation phase, whereas $\eta = -0.5$ demonstrates the softest stress response for all crystal orientations. For the $\eta = 0$ loading scenario, the stress response increases monotonically in all orientations. Figure~\ref{fig:inplanestressratioplot.png}a shows that the stress biaxiality ratio is greater than 0 (positive) for $\beta$ = -0.5, 0 and 1, and $\eta$ =0.8 and 1 loading cases. As a result, in the initial deformation stage, a stiffer stress response is observed, followed by a softening response due to significant void expansion in the crystal, which cannot be further compensated by an increase of average stress in the bulk crystal.
%As a result, in the initial deformation phase, a stiffer stress response is observed, followed by a softening response due to rapid void expansion in the crystal. 
When the magnitude of peak stress for the different orientations is compared, orientation C has the largest peak stress, and orientation A has the lowest peak stress for $\beta$ =1 loading case. Furthermore, the evolution of the overall mean stress in Fig. 7c correlates well with the displacement biaxiality ratio $\beta$ shown in Fig. 7b. In particular, the higher $\beta$ value the more stiff the initial response is and the sooner (in terms of the value of $F_{22}-1$) the peak stress is achieved for the given process.
%Furthermore, the overall mean stress response correlates well with the displacement biaxiality ratio plots (refer Fig \ref{fig:inplanestressratioplot.png}b).

Figure \ref{fig:inplanemeanstressori.png} depicts the overall mean stress response in both pristine and porous unit cells for various crystal orientations and the specified loading scenario. Five loading scenarios where $\beta$ and $\eta$ are both equal to 0 and 1 are analysed, together with scenario $\eta=0.8$ which approximately corresponds to $\beta=0$ case as discussed before. It is evident that all loading cases exhibit the anisotropic response. Also, it is apparent that the response of the porous crystal differs substantially from that of the pristine crystal. With the exception of $\eta= 0$ (uni-axial loading condition), the pristine crystal response is purely elastic. In the case of $\eta = 0$ loading, the pristine crystal displays a stiffer response than the porous crystal for orientations A and C; however, for orientation B, the response is almost the same for both the pristine and porous crystal. The response of orientation C is the stiffest in each of the loading conditions. For loading scenarios, $\beta = 0, 1$, and $\eta = 0.8, 1$, orientation A initially displays the softest response, whereas orientation B exhibits the softest response by the end of the deformation process. When the loading scenarios $\beta = 0$ and $\eta = 0$ are compared, the substantially higher stress biaxiality in the $\beta = 0$ loading case (refer to Fig. \ref{fig:inplanestressratioplot.png}a) causes a more stiff response during an early deformation stage, followed by a softening due to significant void expansion in the crystal. On the contrary, a monotonic stress increase is observed for the $\eta = 0$ loading scenario. Instead, as expected, the stress response for $\beta = 0$ case is close to  $\eta=0.8$ loading conditions. Due to the highest stress biaxiality, a similar response was observed for $\beta = \eta = 1$.

\paragraph{Normalized void volume fraction evolution}
Figure \ref{fig:inplanestressratioplot.png}d compares the evolution of the normalized void volume fraction for various loading conditions and the specified orientation. These evolution plots are in good agreement with displacement biaxiality ratio plots in Fig. \ref{fig:inplanestressratioplot.png}a. For all orientations, the void growth rate increases as the displacement biaxiality ratio increases. The void is collapsing for the $\eta =-0.5$ loading case, and this behaviour is the most pronounced in orientation C. Confirmation of the phenomenon can be seen in contour plots of accumulated shear for orientation C which are shown in Fig. \ref{fig:oriC_accshear_0.3_Rev1.png}. Under $\eta = 0$ loading case, the void grows very slowly as compared to higher stress biaxiality cases. If the displacement biaxiality ratio is less than -0.5, softening behaviour is not observed, since not much void growth is seen. The void growth increases at first with $\beta=-0.5$ but subsequently stabilizes for all orientations. The evolution of the void volume fraction under the $\beta$ and $\eta = 1$ loading scenarios correlates with the evolution of the displacement biaxiality ratio (Fig \ref{fig:inplanestressratioplot.png}b). The curves of void evolution start to deviate from each other at the same moment when the value of $\beta$ for $\eta = 1$ case drops below one.

Qualitative differences are observed in the curves shown in Fig. \ref{fig:inplanestressratioplot.png}d for $\eta$-cases, which can be explained by the accompanied variation of displacement biaxility ratio. As it is seen, for the high stress biaxiality ratio: $\eta=1$, initially for all three orientations the displacement biaxiality $\beta$ is equal to 1, so the cell is under the conditions beneficial for the void expansion. Accordingly, at this stage, the void is growing in all directions (compare Fig. S.2f, S.3f, and S.4f in the supplementary material). However, as deformation proceeds \emph{the displacement biaxiality ratio is \textbf{decreasing towards zero}}, which effectively \emph{\textbf{slows down} the void growth} since its growth starts to be limited to one direction in the plane. Nevertheless, the void volume fraction is still growing on the cost of bulk crystal, and the achieved values are high. This causes a decrease of the overall in-plane mean stress as an increase of average stress in the bulk crystal is not able to compensate for the void expansion, Fig. \ref{fig:inplanestressratioplot.png}c. On the contrary, for smaller stress biaxiality ratio: $\eta=0$, the initial displacement biaxiality is negative, so even though the net change of void volume fraction is positive, in this scenario the void diameter is growing only in one direction while decreasing in the perpendicular one (compare Fig. S.2e, S.3e and S.4e in the supplementary material). For this case, as the deformation proceeds \emph{the displacement biaxiality ratio \textbf{increases towards zero}}, which in this case leads to \emph{\textbf{accelerated} void growth} because the reduction of void size in one of the directions is halted, while it is still growing in another one. For all three orientations for the considered deformation range, the increase in the void volume fraction is not yet sufficient to overcome the overall mean stress increase due to the strain hardening in bulk crystal. However, with increasing deformation one may expect softening which will be accompanied by an accelerated void growth rate.  Interestingly, for the $\eta=0.8$ case the former and latter scenarios of void growth are observed for orientations C and B, respectively (see also Fig. \ref{fig:normvoidoriplot.png}c). Orientation A exhibits here some limit case with the almost constant rate of void volume fraction.

 Figure \ref{fig:normvoidoriplot.png} compares the evolution of the void volume fraction for different crystal orientations and the selected loading conditions. Five loading cases, the same as in mean stress response plots shown in Fig. \ref{fig:inplanemeanstressori.png}  are illustrated. For $\eta= 0$ (in-plane uniaxial loading case), the anisotropic response is observed. Void growth in orientation C is the highest, followed by orientations A and B. But for $\beta=0$ and 1 loading cases, due to relatively large strain biaxiality, the void growth is significant and the effect of crystal orientation diminishes. It has been verified that the latter observation is true also for other processes in which $\beta$ value is kept constant.
 %But for $\beta=0$ and 1 loading case, due to higher stress biaxiality the void growth is rapid and the effect of crystal orientation diminishes.
 The same observation is reported in \citep{POTIRNICHE2006921} under displacement controlled boundary conditions. The void growth for orientations C and B are nearly identical for $\eta = 1$ loading. However, the void growth rate is slower in asymmetric orientation A than in orientations B and C. For the same $\eta$ and $\beta$ values, the void expansion under $\beta = 0$ is substantially faster than the $\eta = 0$ loading situation, since, as already mentioned, this case corresponds approximately to $\eta=0.8$ case, so a much higher stress biaxiality ratio. Similarity between $\eta=0.8$ and $\beta=0$ case is also seen when comparing the contour plots in (b) parts of figures in Subsection \ref{Sec:local} with respective maps in Fig. S.5 in supplementary material.
 %This is because the stress biaxiality is higher in the $\beta= 0$ loading condition.

Unlike in \citep{SubrahmanyaPrasad2015NumericalSO} the void coalescence criterion is not formulated in the present study. Nevertheless, in order to closely observe this phenomenon, Fig. \ref{fig:coalescenceplot.png} it is demonstrated how the void size is changing in three different directions: AB, EF, E'F' marked in Fig. \ref{unitcellmodel} for two selected loading cases: $\beta=1$ and $\eta=1$. The figure presents the evolution of the value of $\log(L/|\mathbf{x}_{\rm{right}}-\mathbf{x}_{\rm{left}}|)$ where $\mathbf{x}_{\rm{right}}$ and $\mathbf{x}_{\rm{left}}$ denote current locations of nodes at the right and left end of the respective diameter and $L$ is the current cell size in a relevant direction. When this value is tending to zero the coalescence is approached. It is seen that for the case $\beta=1$ and symmetric orientations B and C  the coalescence state is attained in two perpendicular bands along X and Y directions. Additionally, the void is loosing its spherical shape, more importantly for orientation B than C. For other cases the coalescence is approached mainly in X direction and this state is attained visibly sooner for orientations A and B than for orientation C. Figure \ref{fig:coalescenceplot.png}a shows that, although the orientation effect is not seen in the normalized void evolution plots shown in Fig. \ref{fig:normvoidoriplot.png} under the same value of displacement biaxiality, it manifests in the developed void shape and thus may influence the coalescence strain and, in general, the failure mode.

\subsubsection{Local sample response\label{Sec:local}}

\begin{figure*}[h!]
    \centering
    \includegraphics[angle=0,width=\textwidth]{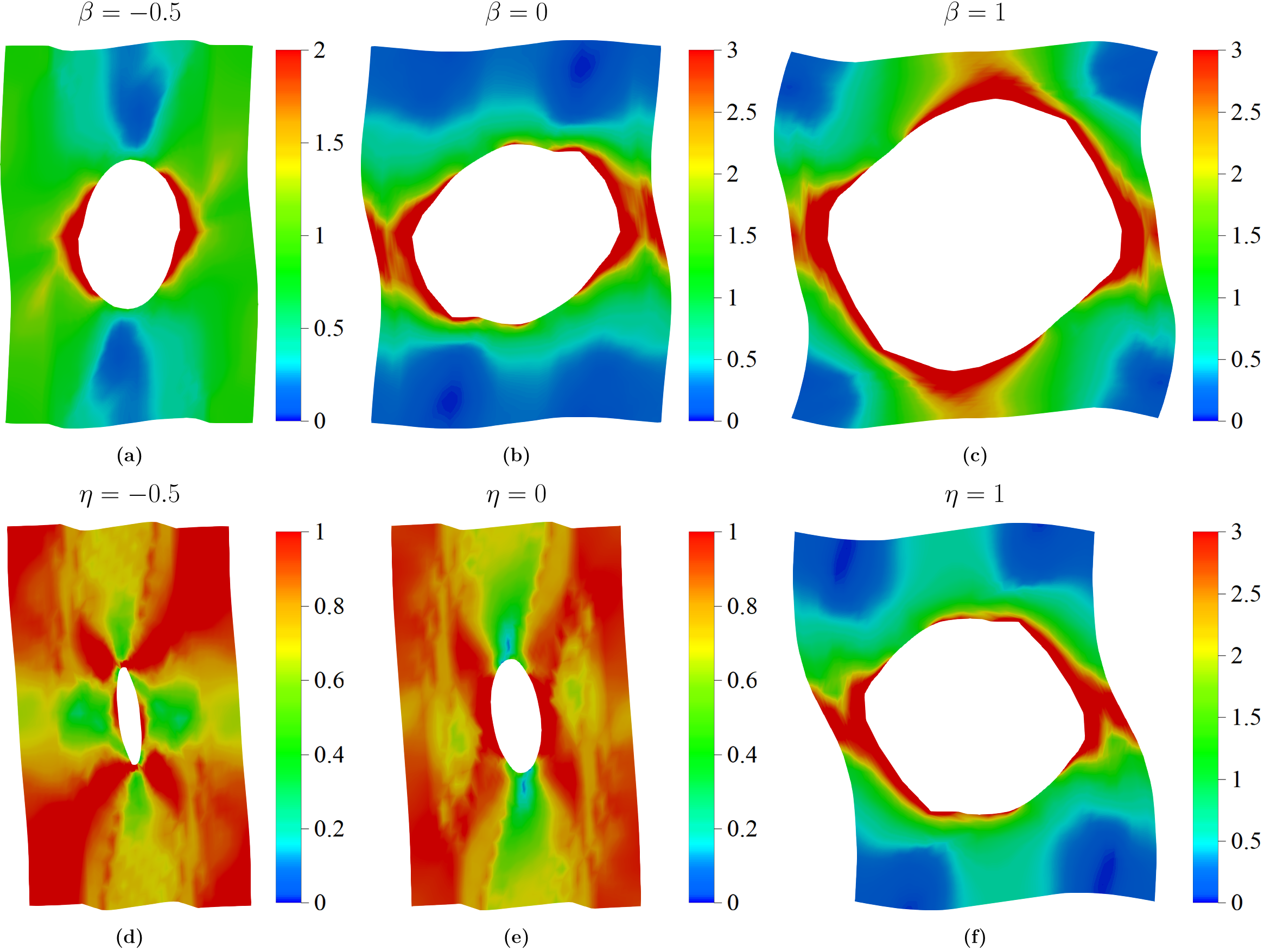}
    \caption{Contour plots of accumulated shear $\Gamma$ for the asymmetric  orientation A under various loading conditions at the strain level of $F_{22}-1 = 0.3$. }
    \label{fig:oriA_accshear_0.3_Rev1.png}
\end{figure*}

\begin{figure*}[t!]
    \centering
    \includegraphics[angle=0,width=\textwidth]{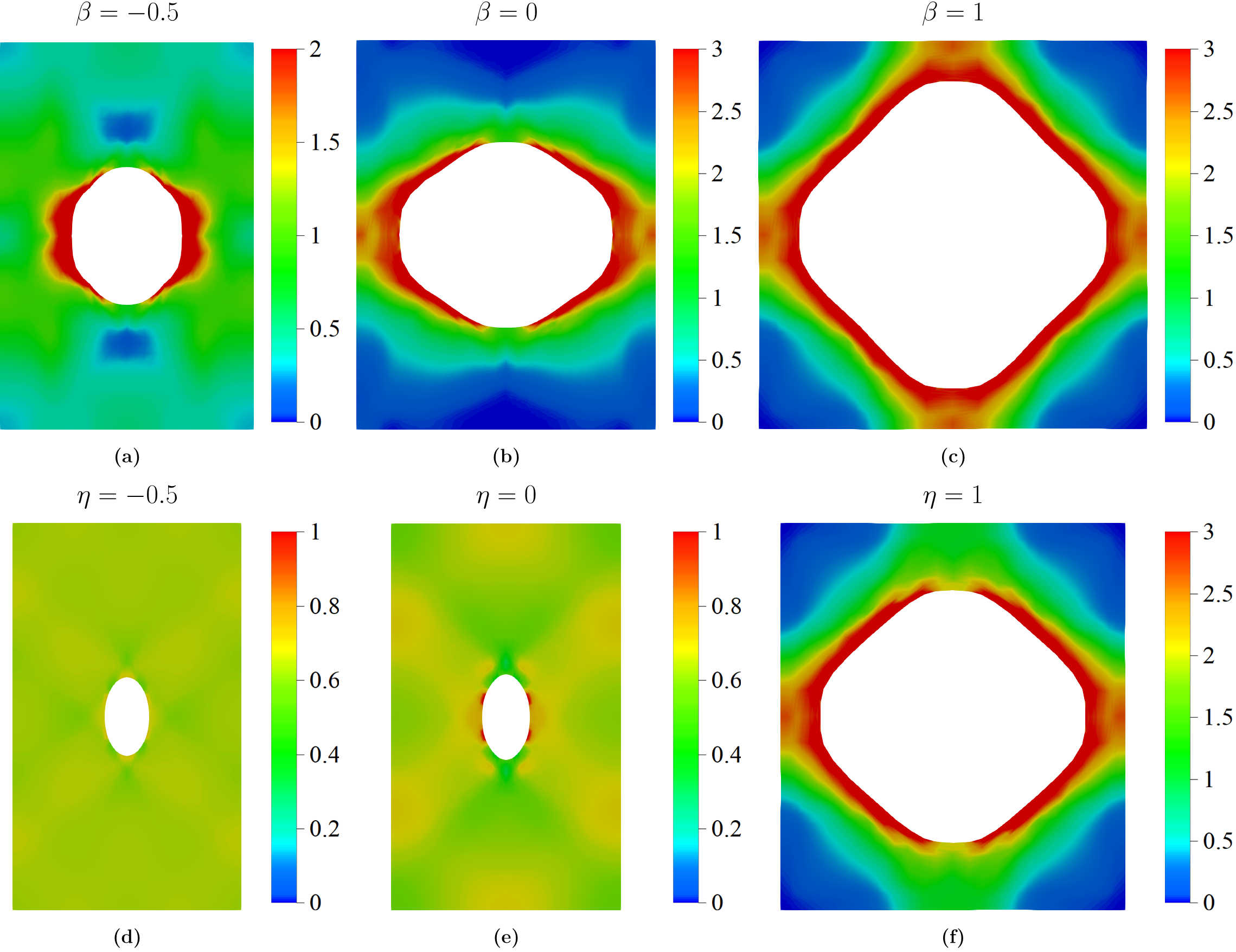}
    \caption{Contour plots of accumulated shear $\Gamma$ for the symmetric  orientation B under various loading conditions at the strain level of $F_{22}-1 = 0.3$. }
    \label{fig:oriB_accshear_0.3_Rev1.png}
\end{figure*}

\begin{figure*}[t!]
    \centering
    \includegraphics[angle=0,width=\textwidth]{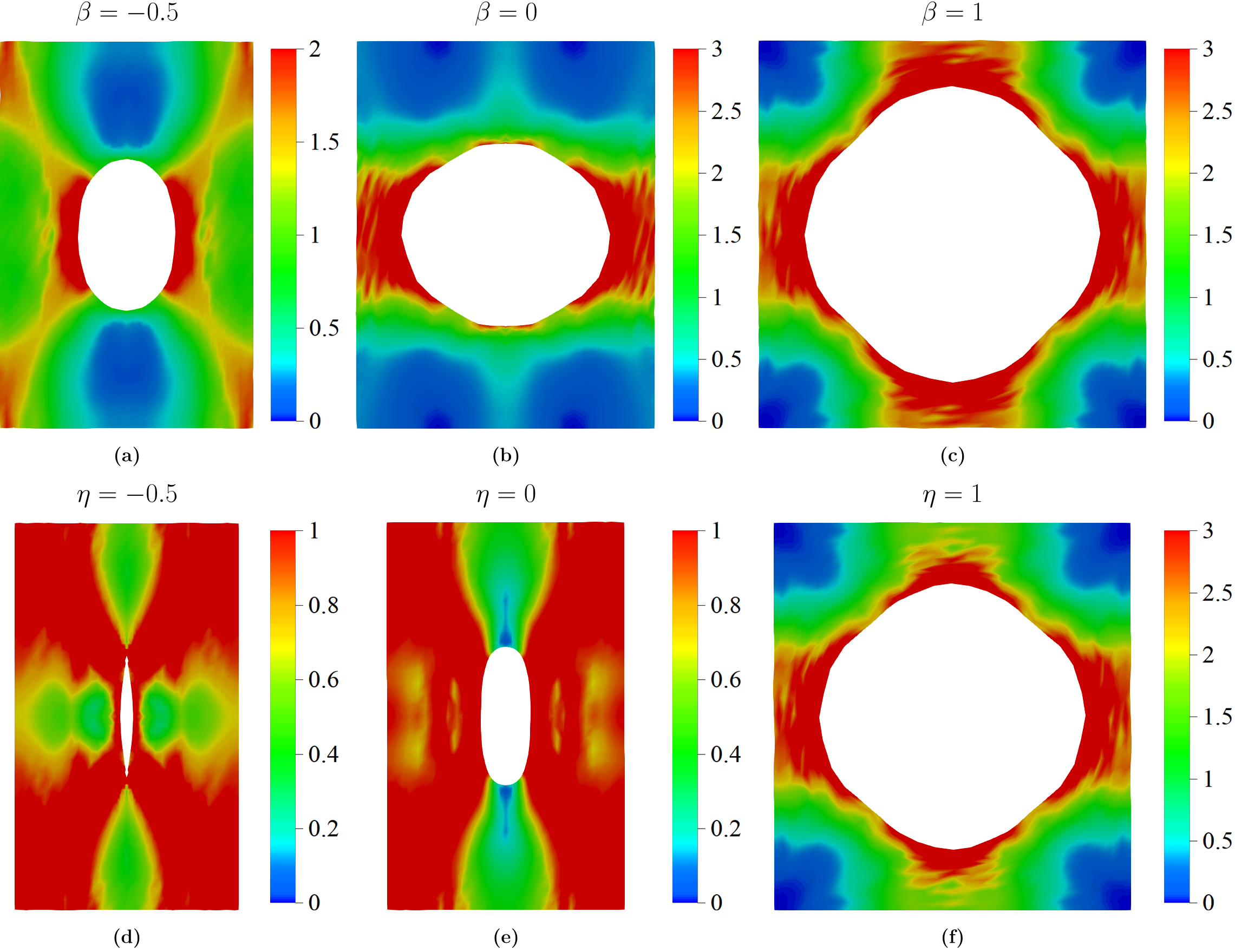}
    \caption{Contour plots of accumulated shear $\Gamma$ for the symmetric  orientation C under various loading conditions at the strain level of $F_{22}-1= 0.3$. }
    \label{fig:oriC_accshear_0.3_Rev1.png}
\end{figure*}

\paragraph{Local distribution of accumulated shear}
{First, in order to show the possible failure mode, contour plots of accumulated shear are presented at}  the end of the deformation process at strain level 0.3 for six considered loading scenarios in Figs. \ref{fig:oriA_accshear_0.3_Rev1.png}, \ref{fig:oriB_accshear_0.3_Rev1.png} and \ref{fig:oriC_accshear_0.3_Rev1.png}, for  three crystallographic orientations A, B, and C listed in Table \ref{tab:crystal orientations}, respectively. Since the strain level along the principal loading direction is the same for all the cases, one is able to observe relative variation in a shape change of the cell as a whole and the void for all six loading scenarios. Additionally, to illustrate the strain localization process, the contour plots are presented for $F_{22}-1 = 0.15$, so at the intermediate stage of the deformation, and placed in the supplementary material.

Part (a) of figures \ref{fig:oriA_accshear_0.3_Rev1.png}-\ref{fig:oriC_accshear_0.3_Rev1.png} shows the contour plots of accumulated shear under $\beta = -0.5$ loading. For all orientations, shear begins to accumulate on the transverse sides of the voids at the intermediate strain level of 0.15. Due to the symmetry of crystal orientations B and C with respect to the loading direction, the symmetrical distribution of accumulated slip is observed, whereas for asymmetrical orientation A alternate bands of severe deformation and no deformation are developing. Because of the relatively high stress biaxiality ratio in the $\beta = -0.5$ scenario, void growth is rapid as deformation progresses (refer to Figs. \ref{fig:inplanestressratioplot.png}a and \ref{fig:inplanestressratioplot.png}d).
The unit cell is deformed substantially at strain level 0.3, with the maximum shear accumulating on the transverse sides of the void. For orientations A and B, the transverse ligament is the origin of void coalescence. In orientation A, the void rotates, and the strain concentration is observed on the transverse sides along the inclined direction. Moreover, at the strain level 0.3 a slight trace of the shear band is seen. Because of the asymmetric orientation, the unit cell edges do not remain straight and {are deformed}. For orientation B, a polygonal void shape is noticed. For orientation C, inclined shear bands clearly form, and the shear accumulates along the transverse sides of the void. The mode of failure in this case is through these inclined shear bands. In addition, the void elongates in the loading direction, resulting in an ellipsoidal void shape.

Part (b) of Figs. \ref{fig:oriA_accshear_0.3_Rev1.png}-\ref{fig:oriC_accshear_0.3_Rev1.png} displays the contour plots of accumulated shear under $\beta=0$ loading. The void growth is substantially faster due to the high stress biaxiality ($0.5<\eta<1$, refer to fig \ref{fig:inplanestressratioplot.png}a) and is evident even at the intermediate strain level of 0.15. At this strain level, shear begins to accumulate around the void and the void starts to grow significantly in the transverse direction for all three orientations. Additionally, the void rotates for orientation A. More shear is accumulated in the transverse ligament for orientation C than for orientations A and B. The symmetric distribution of contours is found again for symmetric orientations B and C. 
As the deformation process proceeds, rapid void growth is observed in all crystal orientations, and coalescence occurs along the transverse sides of the void. The void is substantially rotated for orientation A, and a zigzag pattern of strain localization bands is seen along the transverse direction of the void. Similarly, in orientation B, void expansion in the transverse direction is quick, and a polygonal form of the void is clearly developed. On the other hand, the void shape in orientation C is nearly circular, and accumulated shear is seen in the transverse ligament.

Part (c) of figures \ref{fig:oriA_accshear_0.3_Rev1.png}-\ref{fig:oriC_accshear_0.3_Rev1.png} present the accumulated shear distribution under $\beta = 1$ loading. The contour plots resemble those from the preceding loading scenario, i.e. $\beta$ = 0, however, the void growth is quick in both longitudinal and transverse directions due to the strong stress biaxiality. When compared to the previous loading instance $\beta$ =0, the void expansion and accumulation of shear is significantly more severe at the intermediate strain level of 0.15. The void is rotated for asymmetric orientation  A, as in prior loading scenarios. Due to the high stress biaxiality, the void form is much more circular for orientations B and C at the strain of 0.15. In contrast to prior loading examples, coalescence is observed in both directions at the final strain level of 0.3 for orientations B and C. In addition, for orientations B and C, a polygonal void shape with rounded corners is observed. Similar behaviour was reported in ~\citep{SRIVASTAVA201510} at high stress triaxialities. Furthermore, for both orientations, substantial shear accumulation is seen around the void. The void rotates even further in asymmetric orientation A, but its shape is not perfectly circular or polygonal. The same rapid void growth is clearly observed in normalized void volume fractions plots for this $\beta$ loading case and three orientations (refer Fig.~\ref{fig:inplanestressratioplot.png}d).

Now, let us move to the $\eta=\rm{const}$ loading scenarios.

Subfigures (d) of figures \ref{fig:oriA_accshear_0.3_Rev1.png}-\ref{fig:oriC_accshear_0.3_Rev1.png} show the contour plots of accumulated shear under the $\eta=-0.5$ loading scenario. In comparison to the $\beta= -0.5$ loading condition, void growth is not significant under this loading configuration. This is because the stress biaxiality ratio is low. Also, as previously discussed on the basis of the void evolution plots, void expansion is not detected when the displacement biaxiality ratio $\beta$ is lower than -0.5. At the intermediate strain level of 0.15, inclined shear bands begin to form for orientation A, whereas shear bands for orientation C are at 45 degrees with the main loading direction. However, in orientation B, the deformation is almost homogeneous and there is no void growth. 
For orientations A and C, the void collapses at a strain level of 0.3. Normalized void volume fraction plots confirm the observation. The void in orientation C is collapsing like a penny shaped crack. Furthermore, shear accumulates the most at the tip of the severely elongated void. For orientation B, still, almost homogeneous deformation is seen, with no void expansion.

Part (e) of Fig \ref{fig:oriA_accshear_0.3_Rev1.png}-\ref{fig:oriC_accshear_0.3_Rev1.png} displays the contour plots of accumulated shear in accordance with the $\eta=0$ (uniaxial) loading scenario. The response is quite similar to the loading case with $\eta=-0.5$. At the strain level of 0.15, for asymmetric orientation A, the shear starts to accumulate on the transverse sides of the void and the formation of one family of inclined shear bands is observed whereas for orientation B almost homogeneous deformation is observed with no shear localization. The formation of two families of the inclined deformation bands and accumulation of slip on transverse sides of the void is seen for orientation C.  For orientation A, a noticeable formation of another family of inclined deformation bands is observed at a strain level of 0.3, and the unit cell is distorted whereas for orientation B some heterogeneity of deformation is observed, but not much void expansion. For orientation C, slip shear accumulates on both sides of the void, causing the void to elongate along the loading direction, resulting in an ellipsoidal shape. When compared to orientations A and B, void expansion is substantially more prominent for orientation C. Overall, when comparing responses of three orientations with the respective $\beta = 0$ loading case, the void growth is not {significant} due to low stress and negative displacement biaxiality ratios.
On the other hand, as already discussed, the $\beta=0$ case is approximately equivalent to the $\eta=0.8$ case for which the respective accumulated shear maps are shown in Fig. S.5a of the supplementary material. Those contour plots are very similar to the maps shown in (b) subfigures included in this subsection.

Finally, part (f) of figures \ref{fig:oriA_accshear_0.3_Rev1.png}-\ref{fig:oriC_accshear_0.3_Rev1.png} depicts the accumulated shear contour plots under the $\eta=1$ loading scenario. Similarly to the $\beta=1$ loading scenario, void growth is accelerated due to high stress and displacement biaxiality. The displacement biaxiality ratio plot (Fig. \ref{fig:inplanestressratioplot.png}b) explains the slight deviations from the $\beta=1$ loading condition. For orientations B and C, the void growth is rapid even at the strain level of 0.15, which is identical to the $\beta=1$ loading situation. However, there is some deviation in the strain accumulation as compared to $\beta=1$ case for asymmetric orientation A, which can be correlated with differences seen in the displacement biaxiality ratio curves in Fig. \ref{fig:inplanestressratioplot.png}b for these cases. 
A polygonal void shape with rounded corners is seen for orientations B and C at the strain level of 0.3, which is identical to the $\beta=1$ loading situation. The void growth is slightly reduced since the displacement biaxiality ratio is less than 1 (refer to Fig. \ref{fig:inplanestressratioplot.png}d). The primary difference is that in the $\beta=1$ loading situation, void coalescence occurs in both loading directions, but in the $\eta = 1$ loading instance, void coalescence happens only in the transverse ligament. Displacement biaxiality plots (Fig. \ref{fig:inplanestressratioplot.png}b) clearly show the origin of this disparity. In addition, the maximum shear accumulates around the void for all orientations.

\paragraph{Local distribution of lattice rotation}
The influence of void evolution and loading conditions on new grain formation is now studied on the basis of contour plots of the lattice rotation angle. We concentrate on, somewhat opposite, cases of orientation A and B, and present only selected results for orientation C. 

The lattice rotation angle $\Psi\in (0,\pi)$, presented in the plots, is defined as:
\begin{equation}\label{Eq:miorient-angle}
    \Psi = \arccos{\left(\frac{\rm{tr}(\Delta \mathbf{R}(t)) - 1}{2}\right)}
\end{equation}
where $\Delta\mathbf{R}(t)$ is calculated based on the initial orientation tensor $\mathbf{R}(0)$ and current orientation tensor $\mathbf{R}(t)$, respectively, as
\begin{equation}
     \Delta \mathbf{R}(t) = \mathbf{R}(t)\mathbf{R}(0)^\intercal\,.
\end{equation}
Orientation tensor $\mathbf{R}(t)$ is constructed based on the current orientation of lattice direction $\mathbf{a}$ with the Miller indices $[100]$ and lattice plane normal $\mathbf{b}$ with $\{001\}$, respectively. The change of their orientation during the deformation process is governed by the elastic part of the deformation gradient $\mathbf{F}_e$, so that $\mathbf{a}(t)=\mathbf{F}_e(t)\mathbf{a}(0)$ and $\mathbf{b}(t)=\mathbf{F}_e^{-T}(t)\mathbf{b}(0)$. 

%% Asymmetric orientation A 

\begin{figure*}[h!]
    \centering
    \includegraphics[angle=0,width=\textwidth]{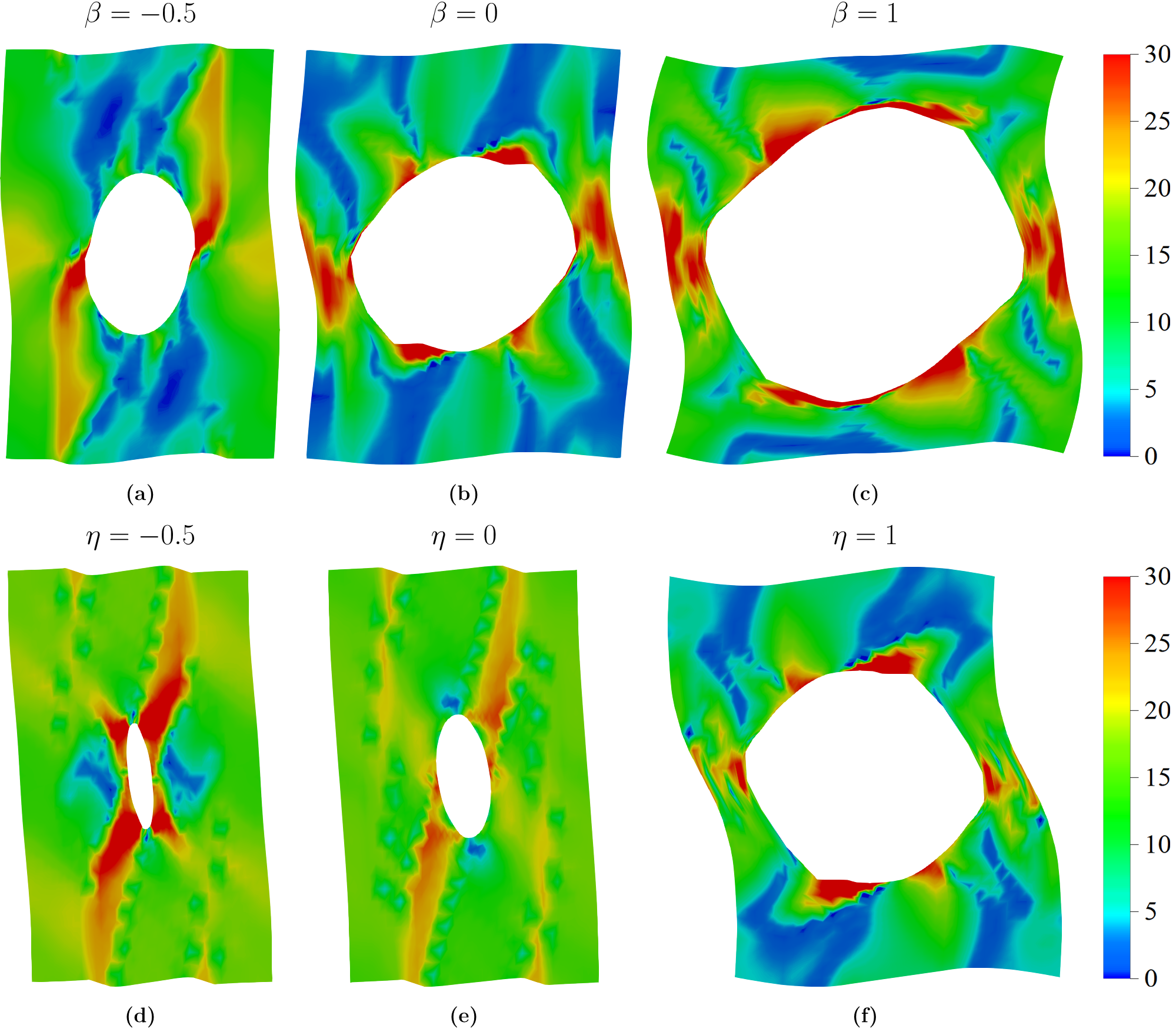}
    \caption{Contour plots of lattice rotation angle ($\Psi$) in (\textbf{degrees}) for the asymmetric  orientation A at the strain level of $F_{22}-1$ = 0.3. }
    \label{fig:Rotang_oriA_0.3.png}
\end{figure*}

%% orientation B

\begin{figure*}[t!]
    \centering
    \includegraphics[angle=0,width=\textwidth]{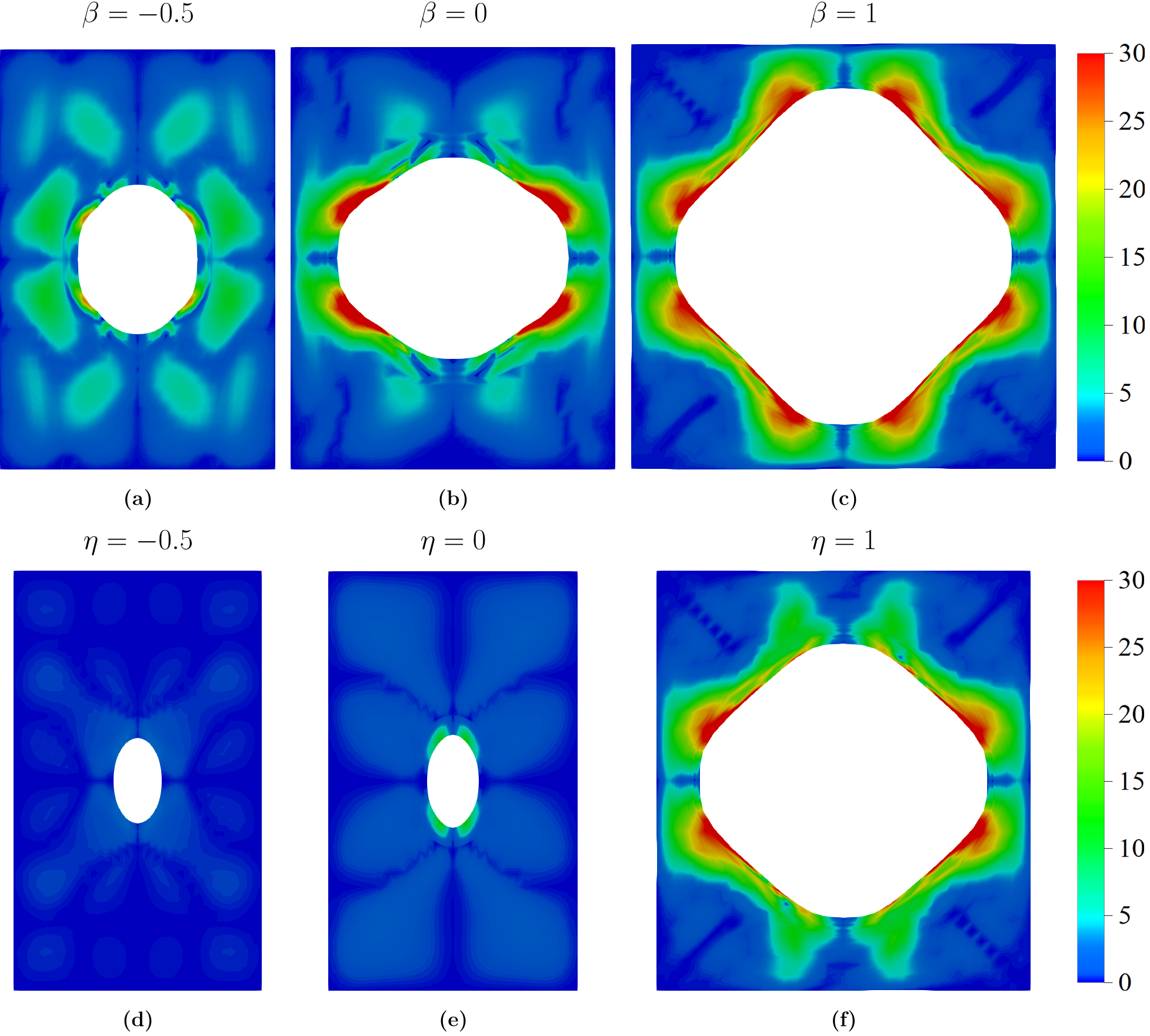}
    \caption{Contour plots of lattice rotation angle ($\Psi$) in (\textbf{degrees}) for the symmetric  orientation B at the strain level of $F_{22}-1$ = 0.3. }
    \label{fig:Rotang_oriB_0.3.png}
\end{figure*}

In each loading case, we observe rotation angle heterogeneity, which results in the development of a new microstructure. The presence of a void causes heterogeneity of strain, which results in heterogeneity of lattice rotation. However, we notice that the distribution of the rotation angle does not follow perfectly the distribution of accumulated shear $\Gamma$, as was already seen in Section~4.1 for uniaxial loading cases.

The lattice rotation angle plots for asymmetric orientation A are shown in Figure \ref{fig:Rotang_oriA_0.3.png}.  Because orientation A is not stable under prescribed loading conditions, we {observe} uniform lattice rotation even in pristine crystals.  For example, the calculated lattice rotation angle for loading case $\eta=0$ at $F_{22}-1=0.3$ for a crystal without a void is $11\degr$. For voided crystal and $\eta=0$ case, we observe bands with rotation angles of $30\degr$, whereas the remaining portion of the cell rotates at a smaller angle of about $10\degr$, which roughly corresponds to the lattice rotation which would be seen in pristine crystals. In the unit cell, new grains are formed as a result of the different rotation angles. Under the $\eta$ = -0.5 loading scenario, a similar response is observed. Two inclined bands are forming in this case. The evolution of the void volume fraction has an effect on the evolution of the microstructure. For $\eta$ = -0.5, we see that the new subgrains with larger rotation angle correlate with the zones of increased accumulation of shear. For $\beta$ =0, $\beta$ =1, and $\eta$ = 1 loading cases, the formation of new grains takes place around the void, as well as alternate domains of no rotation (blue domain) and moderate rotation outside of the void is observed. All of these factors contribute to the formation of multigrain microstructures, particularly at high strain or stress biaxiality values. For the $\beta$ = -0.5 loading case (and for the approximately equivalent $\eta=0.8$ case as seen in the supplementary material), the combination of effects found for other loading cases is seen. We observe the formation of a band with high lattice rotation, which starts at the lateral sides of the void and then is parallel to the main direction of loading, as well as alternating bands of no and medium rotation angles in the middle vertical portion of the unit cell.

Figure \ref{fig:Rotang_oriB_0.3.png} depicts the lattice rotation angle plots for orientation B. Due to the symmetry of orientation with respect to the loading conditions, the developed microstructure preserves this symmetry. Because orientation B is stable under prescribed loading conditions, we do not  observe lattice rotation for pristine crystal. For a voided crystal, for $\eta=0$ and $\eta=-0.5$ loading cases, the heterogeneity of lattice rotation angle is very small, which is less than $5\degr$, following homogeneity of deformation seen in Fig. \ref{fig:oriB_accshear_0.3_Rev1.png}(d, e). Around the void, a few small domains with $5 - 10\degr$ of lattice rotation are present. The formation of new grains around the void is observed for higher biaxiality loading cases where $\beta= 0$, 1 and $\eta = 1$, and the crystal domain in a larger distance from the void does not rotate significantly. Under $\beta= -0.5$ and $\eta=0.8$ loadings, the response is somewhere in between the two scenarios discussed before. 

\begin{figure*}[h!]
    \centering
    \includegraphics[angle=0,width=\textwidth]{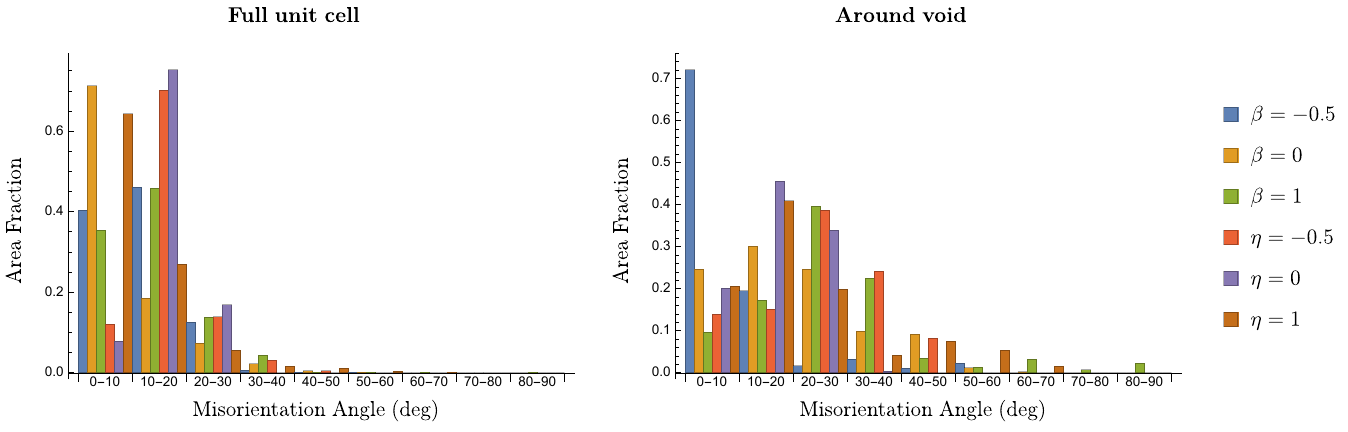}
    \caption{Histogram plots of lattice rotation angle $\Psi$ in \textbf{degrees} for the asymmetric orientation A at the strain level of $F_{22}-1$ = 0.3. (Area fraction is calculated in reference configuration)}
    \label{fig:historiA.pdf}
\end{figure*}

\begin{figure*}[h!]
    \centering
    \includegraphics[angle=0,width=\textwidth]{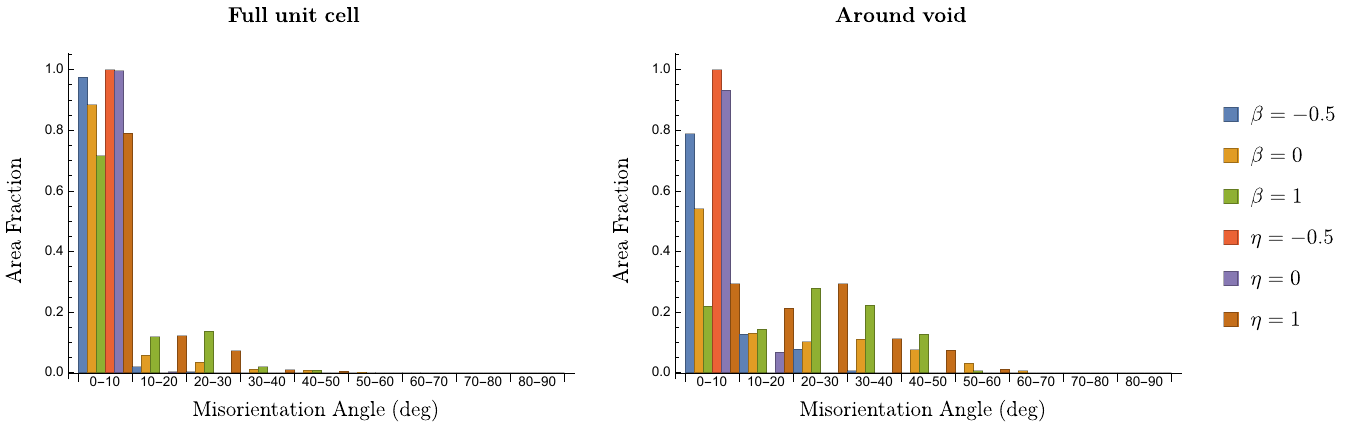}
    \caption{Histogram plots of lattice rotation angle $\Psi$ in \textbf{degrees} for the symmetric orientation B at the strain level of $F_{22}-1$ = 0.3. (Area fraction is calculated in reference configuration)}
    \label{fig:historiB.pdf}
\end{figure*}

\begin{table}[!t]

    \centering
    \begin{tabular}{|l|l|c|c|c|c|c|c|c|}
    \hline
    \multicolumn{2}{|c|}{Loading Case} & $\beta = -0.5$ & $\beta = 0$ & $\beta = 1$ & $\eta = -0.5$ & $\eta = 0$& $\eta = 0.8$ & $\eta = 1$\\
    \hline
    \multirow{2}{*}{Orientation A} & M & 11.656 & 8.593 & 13.605 & 16.804 & 16.034 &9.1274 & 9.308\\
    \cline{2-9}
    & SD & 7.302 & 7.997 & 9.114 & 6.390 & 4.569 &6.91265 & 8.223\\
    \hline
     \multirow{2}{*}{Orientation B} & M & 3.179 & 4.446 & 7.548 & 0.255 & 0.808 &4.08305 & 5.702\\
    \cline{2-9}
    & SD & 2.938 & 7.605 & 9.837 & 0.321 & 0.956 &6.87856& 8.125\\
    \hline
    \multirow{2}{*}{Orientation C} & M & 5.188 & 4.490 & 6.502 & 7.103 & 6.445 &4.68702 & 5.727\\
    \cline{2-9}
    & SD & 5.278 & 6.808 & 7.576 & 8.595 & 6.451 &7.06543& 6.737\\
    \hline
    \end{tabular}
    \caption{Mean (M) and standard deviation (SD) of misorientation angles (degrees) for two orientations under different loading scenarios}
    \label{tab:histograms}
\end{table}

In order to further illustrate the effects related to grain refinement,
Figures \ref{fig:historiA.pdf} and \ref{fig:historiB.pdf} present histograms of the lattice rotation angle generated based on the data in Figures \ref{fig:Rotang_oriA_0.3.png} and \ref{fig:Rotang_oriB_0.3.png}, respectively. The histogram plot on the left displays the entire unit cell, while the histogram plot on the right shows the area surrounding the void.  For the purpose of the latter plot, we employed two layers of finite elements which surround the void (refer to Fig. \ref{fig:mesh}). Different colours of bars correspond to different loading conditions. When we compare the results for asymmetric orientation A and symmetric orientation B, we observe that orientation A has a substantially larger orientation spread than orientation B. This is because most of the elements rotate less than 10 degrees in orientation B. When we consider the area around the void for both orientations, the orientation spread widens significantly, especially for $\eta=1$ and $\beta=1$ loading cases.

In order to quantify more directly observed differences for each case mean value and standard deviation of rotation angle were calculated and collected in Table \ref{tab:histograms}, which includes also the respective values found for orientation C. As expected, for all loading conditions the highest mean value is obtained for orientation A, which is connected with lack of symmetry for this configuration. 
Considering the results for the same orientation but different loading conditions, we see that the highest mean misorientation angle is found for the case $\eta = -0.5$ and 0 for orientation A, and for $\beta = 1$ for orientation B and C (refer to Table \ref{tab:histograms}). The standard deviation is used to illustrate lattice rotation heterogeneity. High biaxiality factors $\eta$ and $\beta = 1$ correspondingly showed the highest lattice rotation heterogeneity for orientation A. However, for other stress and strain biaxiality factors, the variation in lattice rotation is not significantly lower. The smallest heterogeneity is observed for the $\eta = 0$ case (around 4.5 degrees). As a result, in the presence of a void, orientation A is prone to grain refinement. For orientations B and C, the disparities in lattice rotation heterogeneity are larger. Again, $\beta$ and $\eta = 1$ had the highest values. However, no heterogeneity is evident for $\eta = -0.5$ and $\eta = 0$ for orientation B, as shown by contour plots (Figure \ref{fig:Rotang_oriB_0.3.png}). This observation is not true for orientation C, which can be correlated with important strain heterogeneity for those two cases seen in Fig. \ref{fig:oriC_accshear_0.3_Rev1.png}. On overall, the magnitude of grain refinement appears to be more influenced by loading conditions in the case of symmetric orientations.

\section{Summary and conclusions}

In this paper, using the crystal plasticity theory combined with the finite element method, we have investigated the effects of initial crystallographic orientation, stress, and displacement controlled loading conditions on the void and microstructure evolution in a 2D plane strain unit cell. Uniaxial and biaxial loading cases have been studied. 

For uniaxial loading cases a special configuration, which enforces an equivalent pattern of plastic deformation in the pristine crystal, has been selected in order to investigate the mutual interactions between the evolving void and the lattice rotation heterogeneity. It has been found that neither macroscopic in-plane stress biaxiality nor displacement/strain biaxiality, are sufficient to fully decide about the void growth, especially when anisotropic materials are considered, and that a significant role in this process is played by microstructure evolution. Fragmentation of bulk crystal surrounding the void into subgrains may lead to significant disturbance of the void volume changes. Note that a similar observation, about the importance of the microstructure changes, was made by \cite{SubrahmanyaPrasad2015NumericalSO} for HCP crystal in which the appearance of domains with new twin related orientation strongly affected void growth and coalescence.

Next, biaxial loading cases have been considered for three crystal orientations, one of which is not symmetric with respect to loading directions. It has been analysed how stress or strain biaxility factors and initial lattice orientation influence the void evolution in terms of its size and shape. Overall, seven cases with three displacement controlled loading scenarios ($\beta=\{-0.5, 0,1\}$) and four stress controlled loading scenarios ($\eta=\{-0.5, 0, 0.8, 1 \}$) have been considered. The following are the key conclusions of the study:
\begin{enumerate}
    \item It seems that the primary driving factor for void growth and coalescence is the displacement biaxiality factor $\beta$. A  clearer correlation is found between variations in displacement biaxiality ratio and normalized void volume fraction evolution plots, as well as the resulting void shape and coalescence pattern. 
    
    \item Softening stress response is evident for large displacement biaxiality factors  when the stress biaxiality ratio $\eta$ increases. The void volume fraction increase in such cases is significant, resulting in void coalescence. The effect of crystal orientation is then diminished. Similar findings were reported in other studies \citep{POTIRNICHE2006921}. The coalescence is observed in both directions for displacement biaxiality $\beta$ =1, but only in the transverse ligament for stress biaxiality $\eta$ =1. For advanced plastic deformation, particularly at high stress and displacement biaxiality $\eta = \beta = 1$, voids evolve into polygonal forms. Similar findings have been reported by \cite{SRIVASTAVA201510}.

   \item  For stress controlled processes the starting point can be described as a biaxial straining process, which under the void growth is approaching an uniaxial straining mode. The way by which the void growth proceeds is governed by the variation of the displacement biaxiality factor $\beta$. When initially $\beta$ is positive the obtained void volume fractions are larger (softening is observed earlier), while the void growth rate will be decreasing when the uniaxial straining mode is approached. On the other hand, when initially $\beta$  is negative then the obtained void volume fractions are smaller (softening is observed later), while the void growth rate will be increasing when the uniaxial straining mode is approached.
    
    \item For lower stress $\eta$ and displacement $\beta$ biaxiality values, an anisotropic response is observed, and the strain-stress response is dependent on crystallographic orientation. For the lowest value of stress biaxiality $\eta = -0.5$, void closure has been observed, particularly in the non-symmetric orientation A and orientation C, as well as the formation of strain localization bands.
    \item In general, the heterogeneity of plastic deformation is the largest for non-symmetric orientation A. This results in lattice rotation heterogeneity and the formation of grain fragmentation in each loading case. For other orientations heterogeneity of lattice rotation is concentrated around the void, especially for higher stress and displacement biaxiality ratios ($\beta =\{ 0, 1 \}$ \& $\eta=\{ 0.8, 1 \}$). On the other hand, for small or negative values of both biaxiality factors, void evolution, and lattice rotation heterogeneity is greatly influenced by initial crystal orientation and substantially differ for the same value of stress and strain biaxiality factor, while the grain refinement encompasses a larger crystal volume.
\end{enumerate}
It should be remarked that FCC crystals usually present smaller plastic anisotropy than HCP crystals, for which different types of slip systems can be activated with substantially different values of critical shear stresses. Moreover, for many HCP metals, uniaxial twinning plays an important role. In such a situation, we may expect even more significant influence of microstructure changes on void evolution and accompanying ductile failure mode. This, together with extending the analysis to 3D spherical voids, is an interesting direction for further research.

\section*{Acknowledgements}
The research was partially supported by project No.  2021/41/B/ST8/03345 of the National Science Centre, Poland. Authors acknowledge Prof. Stanislaw Stupkiewicz from (IPPT) for his help in AceGen/AceFEM implementation of the  computational procedures and Dr. Karol Frydrych (IPPT) for fruitful discussions.

\bibliographystyle{elsart-harv}

\bibliography{refer_IJSS}

\end{document}